\documentclass{svjour3}
\journalname{}

\begin{document}

\title{Can Witten spinor Hamiltonian formulation describe the GR angular momentum?}

\author{Siao-Jing Li}
\institute{Department of Physics, National Central University, Chungli 320, Taiwan, \\
\email{as\_colorful\_as\_a\_chameleon@yahoo.com}}

\date{Received: date / Accepted: date}

\maketitle

\begin{abstract}
The Witten spinor Hamiltonian formulation has previously been shown to be able to yield a quasilocalization for the GR energy-momentum which leads to a proof of the positive energy when the spinor satisfies the Witten equation. In this work we investigate whether this formulation can also describe the GR angular momentum. We conceive four candidates of the quadratic spinor Hamiltonian for the angular momentum based on Witten's scheme. The first one is acquired by substituting the spinor pseudovectorial parameterization for the spinor vectorial parameterization of the displacement in the Witten Hamiltonian. The other three each are composed of other quadratic spinor terms, all with the displacement consisting of the spinor parameterization of an antisymmetric 2-rank tensor and a position vector, one having 4 terms and the others each having twice distinct 2 of the 4 terms. With possible field perturbations around the flat spacetime we find none of the four candidates can give the angular momentum quasilocalization. The importance and prospects for successfully including the angular momentum in Witten's formalism are discussed.

\keywords{Witten spinor Hamiltonian \and quasilocal angular momentum \and gravitation \and spinor-curvature identity \and spinor bivectorial parameterization}
\PACS{04.20.Cv \and 04.20.Fy \and 11.10.Ef \and 11.30.Cp}

\end{abstract}

\section{Introduction}

The principle of equivalence, one of the foundations guiding Einstein to discover his equations of the gravitational field, obstructs the prospect of achieving the tensorial expressions for the densities of 10 conservative quantities (energy-momentum, angular momentum and center-of-mass moment) of an isolated gravitational system though their total values are well defined \cite{MTW}. However, for the gravitational field interacts with matter and other field sources through exchanging their energy-momentums, angular momentums and center-of-mass moments locally, we hope to find some analogue of the densities of the physical quantities of the gravitational field. One of the approaches being developed for this purpose is quasilocal quantities.

There have been many quasilocal proposals \cite{Living}. Various criteria have been addressed to determine good expressions, in particular good limits to the flat spacetime, weak field, spatial infinity (ADM) \cite{ADM} and null infinity (Bondi) \cite{Bondi}. There are still an infinite number of expressions satisfying these requirements. Undoubtedly extra criteria and principles are needed. The Hamiltonian formalism includes such a principle that the Hamiltonian boundary integrals give the quasilocal quantities which depend on the boundary conditions.

The Hamiltonian generating the dynamical evolution along the displacement vector of a gravitational system consists of a spatial 3-volume integral which determines the field equations and a 2-surface boundary integral which is, as argued by Regge and Teitelboim, necessary and can be adjusted to make the Hamiltonian functionally differentiable with respect to the dynamical variables, the latter entailing that the part of the Hamiltonian variation boundary 2-form involved with the dynamical variable variation has a vanishing integral \cite{ReggeTeitelboim,BeigMurchadha,Szabados}. For an asymptotically flat spatial region, when the field equations are satisfied, the volume 3-form vanishes and all the Hamiltonian boundary integrals rendering the well-defined Hamiltonian equations give, for each displacement, the conservative total quantities dependent on the boundary conditions which will be revealed in the symplectic structures of the Hamiltonian variation boundary 2-forms \cite{energyflux,geometricgravity,Quasilocalization,Quasilocalquantities,symplectic,pseudotensorsEM,generalpseudotensors}---particularly, for an asymptotic timelike (spacelike) Poincar\'{e} translation displacement, the boundary integral shall give the total energy (momentum), and for an asymptotic Poincar\'{e} rotation (boost) displacement, the boundary integral shall give the total angular momentum (center-of-mass moment) \cite{Szabados,GravitationalHamiltonian}. It is expected that the integral over a finite 2-surface bounding a 3-region of each Hamiltonian boundary 2-form which renders the Hamiltonian satisfying the above two properties, namely, the involved Hamiltonian variation boundary integral vanishes asymptotically and the Hamiltonian boundary integral approaches the conservative total quantity asymptotically, give the quasilocal quantity of the corresponding conservative quantity \cite{Quasilocalization,Quasilocalquantities}.

Among all Hamiltonians which give the quasilocal quantities, the Witten spinor Hamiltonian \cite{Witten,PositiveEnergy,Anotherpositivity,Newvariables} stands out because besides yielding an energy-momentum quasilocalization, it also gives a proof of the positive gravitational energy, a criterion required by the purely attractive property of gravity, when the spinor satisfies the Witten equation. The Witten Hamiltonian is composed of two quadratic spinor terms which can be, through a spinor-curvature identity \cite{SCI} together with some algebras, transformed into the expected Hamiltonian volume 3-form under the \textit{a priori} torsion-free condition plus a total differential whose superpotential gives an energy-momentum quasilocalization, with a spinor vectorial parameterization for the displacement. However, on the other hand, the prospect of this formalism in yielding a quasilocal expression for the angular momentum keeps not being sufficiently investigated. We expect a full-fledged spinor Hamiltonian formulation can yield quasilocalizations of the 10 conservative quantities for an asymptotically flat gravitational system \cite{SpinorFormulation}. Thus the purpose of the paper is to investigate in detail the feasibility of the spinor formulation for the angular momentum in the regime of the Witten Hamiltonian formalism.

To get the angular momentum, we first consider the Hamiltonian acquired by substituting the spinor pseudovectorial parameterization for the spinor vectorial parameterization of the translation displacement in the original Witten spinor Hamiltonian, but find this parameterization is infeasible for the rotation displacement is a vector rather than a pseudovector. Then we turn back to consider the possibility of the original Witten Hamiltonian for the rotation displacement and find such a displacement will make the Hamiltonian 3-form divergent at the origin. Seeing the hopelessness of the tweak of the original Witten Hamiltonian for the angular momentum, we move on to consider another spinor parameterization. We conceive a 4-term quadratic spinor Hamiltonian whose displacement is made up of the spinor parameterization of an antisymmetric 2-rank tensor and a position vector. Unfortunately, we find this Hamiltonian is unsuccessful for, with possible asymptotic falloffs with parities of the dynamical variables and spinor field, its boundary integrand doesn't approach the standard ADM term asymptotically and the part of its variation boundary integral involved with the dynamical variable variation doesn't vanish asymptotically. From the process of the trial, we speculate the Hamiltonian composed of certain two terms among the 4-term Hamiltonian may promise to be a feasible Hamiltonian for the angular momentum. Therefore we inspect this 2-term Hamiltonian (actually twice the 2-term) and find, with a parity shift among the spinor perturbation, all problems involved in the 4-term Hamiltonian are eliminated but the Hamiltonian boundary 2-form contains a redundant asymptotic term which does not appear in the boundary form of the 4-term one. We also look into the Hamiltonian composed of twice the other 2 terms among the 4-term Hamiltonian with the same parity shift for the spinor and find it has partial problems of the 4-term one along with the redundant asymptotic boundary term which, by a sign difference, appears in the preceding 2-term one.

We expect we can find a certain constraint or introduce some gauge field to cut off the undesired boundary terms without simultaneously taking away the desired terms in either of the latter 3 quadratic spinor Hamiltonians described in the previous paragraph. If we can succeed in finding a Witten Hamiltonian for the angular momentum, we will advance for the exploration of the center-of-mass moment, for which a torsion-related term needs to be added to the Hamiltonian volume term and a M{\o}ller-Komar-like term \cite{Moller,furtherMoller,Komar} to its boundary term. More importantly, with some clues, we expect a successful spinor angular momentum expression will lead us to a proof of a certain connection between the energy and angular momentum of a gravitational system.

We arrange this work as follows. In Sec. \ref{HamiltonianFormulation}, we give the derivation of the Hamiltonian in conventional variables from the Einstein-Cartan Lagrangian. We state the most promising Hamiltonian boundary term choice and lay out how it leads to the conservative total quantities and the quasilocal quantities. We also demonstrate the boundary condition decided by this boundary term from the boundary term of the Hamiltonian variation. In Sec. \ref{Spinor}, we summarize the spinor objects we will employ in our spinor Hamiltonian formulation, including the Dirac matrices, Dirac spinor, and Clifforms. We outline the Dirac spinor parameterizations for several differently-behaving (pseudo-)tensors, the exterior covariant differentials of the spinor, as well as the critical identities, in particular the identities of the Dirac algebra for transforming the volume terms and boundary terms in the Hamiltonians and their variations into the desired patterns along with the generic form of spinor-curvature identities. In Sec. \ref{Pseudovector}, we study the Hamiltonian acquired simply by changing the spinor vectorial parameterization of the displacement of the Witten spinor Hamiltonian into a pseudovectorial one and argue its failure for the angular momentum. In Sec. \ref{4-term}, we explore another conceived 4-term quadratic spinor Hamiltonian, whose displacement includes the spinor parameterization of a bivector and a position vector, for the angular momentum and find it fails. In Sec. \ref{2-term}, we examine the promising 2-term (among the 4-term) spinor Hamiltonian for the angular momentum and find it fails. In Sec. \ref{other2-term}, we inspect the other 2-term (among the 4-term) spinor Hamiltonian for the angular momentum and find it fails. In Sec. \ref{StructuralAnalysis}, we analyze the structure of the quadratic spinor terms in the bivectorial Hamiltonians in the previous three sections by comparing them to the QSL Hamiltonian and its variants. In Sec. \ref{Conclusion}, we conclude our research results and reveal the future work following this work.

\section{The Hamiltonian formulation}\label{HamiltonianFormulation}

For our purpose of introducing the Dirac spinor field, we need to adopt the pseudo-orthonormal frame, so we start with the Einstein-Cartan Lagrangian 4-form \footnote{This is the specific theory adopted in our spinor formulation, being the restricted case of $\mathcal{L}=\textrm{D}g_{\alpha\beta}\wedge\pi^{\alpha\beta}+\textrm{D}\vartheta^\alpha\wedge\tau_\alpha+R^{\alpha\beta}\wedge\rho_{\alpha\beta}-\Lambda(g_{\alpha\beta}, \vartheta^\alpha; \pi^{\mu\nu}, \tau_\alpha, \rho_{\alpha\beta})$.}, which formulates the pseudo-Riemannian General Relativity in terms of the independent variables, the coframe and connection,
\begin{equation}\label{LEC}
\mathcal{L}_{\rm{EC}}=-R^{\alpha\beta}\wedge\rho_{\alpha\beta}+V^{\alpha\beta}\wedge(\rho_{\alpha\beta}-\eta_{\alpha\beta}),
\end{equation}
where $\omega^{\alpha\beta}$ is the connection 1-form in the pseudo-orthonormal coframe $\vartheta^\mu$ in which the metric has the components $\rm{diag}(1,-1,-1,-1)$, $\rho_{\alpha\beta}$ is the canonical momentum conjugate to $\omega^{\alpha\beta}$, $R^{\alpha\beta}=\textrm{d}\omega^{\alpha\beta}+\omega^\alpha{}_\gamma\wedge\omega^{\gamma\beta}$ is the curvature 2-form, $\eta_{\alpha\beta}=\frac{1}{2}\eta_{\alpha\beta\mu\nu}\vartheta^\mu\wedge\vartheta^\nu$ with $\eta_{\alpha\beta\mu\nu}$ being the components of the 4-dimensional spacetime volume form expanded in $\vartheta^\mu$, and $V^{\alpha\beta}$ is a Lagrangian multiplier.
Varying $\mathcal{L}_{\rm{EC}}$, we get
\begin{eqnarray}\label{VLEC}
&&\delta\mathcal{L}_{\rm{EC}}=\textrm{d}(-\delta\omega^{\alpha\beta}\wedge\rho_{\alpha\beta}) \nonumber \\
&&\qquad\quad\>\;-\delta\omega^{\alpha\beta}\wedge \textrm{D}\rho_{\alpha\beta}-(R^{\alpha\beta}-V^{\alpha\beta})\wedge\delta\rho_{\alpha\beta}\nonumber \\
&&\qquad\quad\>\;+\delta V^{\alpha\beta}\wedge(\rho_{\alpha\beta}-\eta_{\alpha\beta})-\delta\vartheta^\mu\wedge V^{\alpha\beta}\wedge\eta_{\alpha\beta\mu}.
\end{eqnarray}
If the integral of the first line of $\delta\mathcal{L}_{\rm{EC}}$, which, through the Stokes theorem, becomes a boundary integral, vanishes, we get the Lagrangian equations \cite{Goldstein}:
\begin{eqnarray}
&&\textrm{D}\rho_{\alpha\beta}=0, \label{vDCM}\\
&&R^{\alpha\beta}=V^{\alpha\beta}, \label{CUeqLA}\\
&&\rho_{\alpha\beta}=\eta_{\alpha\beta}, \label{CMOC} \\
\textrm{and} \qquad &&V^{\alpha\beta}\wedge\eta_{\alpha\beta\mu}=0 \label{vLA}.
\end{eqnarray}
After the combinations of the above four equations, we gain the vanishing torsion and the vacuum gravitational field equations:
\begin{eqnarray}
&&\textrm{D}\eta_{\alpha\beta}=0 , \label{T}  \\
\textrm{and} \qquad &&R^{\alpha\beta}\wedge\eta_{\alpha\beta\mu}=0. \label{R}
\end{eqnarray}

Under a local diffeomorphism along the displacement vector field $N$, $\delta$ being replaced by $\pounds_N$, and inserting (\ref{CMOC}) in (\ref{VLEC}), we obtain the conserved Noether current, the Hamiltonian 3-form,
\begin{equation}\label{H}
\mathcal{H}(N)_{\rm{EC}}=-\pounds_N\omega^{\alpha\beta}\wedge\eta_{\alpha\beta}-\textrm{i}_N\mathcal{L},
\end{equation}
with which $\textrm{d}\mathcal{H}_{\rm{EC}}=0$ is satisfied at the solution of the field equations, (\ref{vDCM}), (\ref{CUeqLA}), (\ref{CMOC}) and (\ref{vLA}). We don't need the information of the Lagrangian multiplier $V^{\alpha\beta}$ anymore, so let us discard it by just substituting (\ref{LEC}) at the solution of the constraint equation introduced by $V^{\alpha\beta}$, that is (\ref{CMOC}), into (\ref{H}) to yield the explicit expression,
\begin{equation}\label{Hamiltonian}
\mathcal{H}(N)_{\rm{EC}}=R^{\alpha\beta}\wedge \textrm{i}_N\eta_{\alpha\beta}+\textrm{i}_N\omega^{\alpha\beta}\textrm{D}\eta_{\alpha\beta}-\textrm{d}(\textrm{i}_N\omega^{\alpha\beta}\eta_{\alpha\beta}).
\end{equation}
When integrated over a spatial 3-volume region with boundary, one can see the first and second terms of (\ref{Hamiltonian}) vanish when the field equations (\ref{T}) and (\ref{R}) are satisfied, leaving the integral of the third term, which becomes a bounding 2-surface integral. However, this boundary term is not uniquely decided---under an addition of a total differential $\textrm{d}\mathcal{B}$ to $\mathcal{H}_{\rm{EC}}$, the Noether current still conserves: $\textrm{d}(\mathcal{H}_{\rm{EC}}+\textrm{d}\mathcal{B})=\textrm{d}\mathcal{H}_{\rm{EC}}=0$. This boundary term needs to be chosen so that the part of the boundary integral of the Hamiltonian variation associated with the variation of the dynamical variables vanishes to guarantee the well-defined Hamiltonian equations.

Regge and Teitelboim argue that by adding a total differential term to the Hamiltonian 3-form, the field equations keep unchanged while different boundary terms correspond to different boundary conditions \cite{ReggeTeitelboim,BeigMurchadha,Szabados}, which are reflected in the covariant symplectic structures of the boundary terms of the Hamiltonian variations \cite{symplectic}, as will be shown later in this section. For an asymptotically flat spatial region, with suitable choices of the asymptotic behavior for the displacement vector fields $N$, the boundary integrals should give conservative total energy-momentum, angular momentum and center-of-mass moment. When the region becomes finite whose boundary is not deviated too much from flatness, the boundary integrals are expected to give the corresponding quasilocal quantities \cite{Szabados,GravitationalHamiltonian}.

By introducing the reference values of the background geometry and selecting the frame gauge $(\pounds_N\vartheta^{[\alpha})^{\beta]}=0$ \footnote{The seeming gauge condition $\pounds_N\vartheta^\alpha=M^\alpha{}_\beta\vartheta^\beta=0$ would, when a (pseudo-)orthonormal frame is adopted, lead to $(\pounds_Ng)_{\alpha\beta}=M_{\alpha\beta}+M_{\beta\alpha}=0$, a tensorial condition. This would not be a problem for the term ${\tilde \textrm{D}}_\beta N^\alpha \textrm{D}\eta_\alpha{}^\beta$ since the symmetric part of $(\pounds_N\vartheta^\alpha)^\beta$ in $(\pounds_N\vartheta^\alpha)^\beta\textrm{D}\eta_{\alpha{}\beta}$ would not make contribution but would be problematical for the term $\stackrel{\circ}
{\tilde\textrm{D}}{}^\beta{\stackrel{\circ}{N}}{}^\alpha\Delta\eta_{\alpha\beta}$ because $\Delta\eta_{\alpha\beta}=\eta_{\alpha\beta}-\stackrel{\circ}{\eta}_{\bar\alpha\bar\beta}e^{\bar\alpha}{}_\alpha e^{\bar\beta}{}_\beta$ is antisymmetric with respect to neither the unbarred indices, $\alpha$ and $\beta$, nor to the barred indices, $\bar\alpha$ and $\bar\beta$. Here we use the barred subscript (e.g., $\bar\alpha$) to denote the components with respect to the background coframe $\stackrel{\circ}{\vartheta}$ and $e^{\bar\alpha}{}_\alpha$ to represent the transformation matrix between the background frame $\stackrel{\circ}{\vartheta}$ and dynamical frame $\vartheta$,  Thus the gauge condition for a (psuedo-)orthonormal frame, which is necessary for introducing the spinor field, begs a special treatment. But $\stackrel{\circ}
{\tilde\textrm{D}}{}^\beta{\stackrel{\circ}{N}}{}^\alpha\Delta\eta_\alpha{}_\beta$ is needed for the center-of-mass moment only and our paper is mainly concerned about the angular momentum, so the special treatment won't be done here.}, which leads to $\textrm{i}_N\omega^\alpha{}_\kappa\equiv\textrm{i}_{\textbf{e}_\kappa}\textrm{i}_N\textrm{D}\vartheta^\alpha+\textrm{D}_\kappa N^\alpha-\textrm{i}_{\textbf{e}_\kappa}\pounds_N\vartheta^\alpha=\Theta^\alpha(N,\textbf{e}_\kappa)+\textrm{D}_\kappa N^\alpha:={\tilde \textrm{D}}_\kappa N^\alpha$ \cite{Quasilocalquantities,symplectic,covariance}, $\textbf{e}$ being the frame dual to $\vartheta$ satisfying $\vartheta^\alpha(\textbf{e}_\beta)=\delta^\alpha{}_\beta$ and $\Theta^\alpha$ being the torsion 2-form, the best Hamiltonian 3-form which is imparted with the most desirable properties for a description of the quasilocal quantities is
\begin{equation}\label{HF}
\mathcal{H}_\vartheta(N)=R^{\alpha\beta}\wedge \textrm{i}_N\eta_{\alpha\beta}+{\tilde \textrm{D}}_\beta N^\alpha \textrm{D}\eta_\alpha{}^\beta+\textrm{d}\mathcal{B}_\vartheta(N),
\end{equation}
with
\begin{equation}\label{B}
\mathcal{B}_\vartheta(N)=-\Delta\omega^{\alpha\beta}\wedge \textrm{i}_N\eta_{\alpha\beta}-\stackrel{\circ}
{\tilde\textrm{D}}_\beta{\stackrel{\circ}{N}}{}^\alpha\Delta\eta_\alpha{}^\beta,
\end{equation}
where $\Delta\omega^{\alpha\beta}:=\omega^{\alpha\beta}-{\stackrel{\circ}{\omega}}{}^{\alpha\beta}$, $\Delta\eta_{\alpha\beta}:=\eta_{\alpha\beta}-\stackrel{\circ}{\eta}_{\alpha\beta}$, with ${\stackrel{\circ}{\omega}}{}^{\alpha\beta}$ and $\stackrel{\circ}{\eta}_{\alpha\beta}$ being the reference values of the connection 1-form and its conjugate momentum, respectively, $\stackrel{\circ}{\textrm{D}}_\beta$ denotes the ground state of the covariant derivative, which is just the exterior derivative if the background geometry is taken as Minkowski space, and ${\stackrel{\circ}{N}}{}^\alpha$ are the reference values of the displacement components. When the field equations are satisfied (on shell), the first term of (\ref{B}) is the ADM term, which makes up the asymptotic forms of the integrands of all conservative quantities for an asymptotically flat spatial region, while the second term with $\stackrel{\circ}
{\tilde\textrm{D}}$ reduced to $\stackrel{\circ}
{\textrm{D}}$ is the M{\o}ller-Komar-like term \cite{Moller,furtherMoller,Komar}, which plays a crucial role in the black hole thermodynamics \cite{geometricgravity,Quasilocalquantities}, certain angular momentum calculations and all center-of-mass moment calculations.

To get the energy-momentum, we prescribe the asymptotic displacement to be the asymptotic Poincar\'{e} translation $N^\mu=\alpha^\mu\approx{\stackrel{\circ}{\alpha}}{}^\mu+{\alpha^{\mu}}^+_1+{\alpha^{\mu}}^-_2$ when expanded in an asymptotic Lorentz frame, where $\alpha^\mu$ are the translation parameters with ${\stackrel{\circ}{\alpha}}{}^\mu$ being its values in Minkowski space, the superscripts $+$ and $-$ designate the even and odd parities, respectively, and $\alpha^\mu{}_s$ denotes the falloff $\rm{O}(\frac{1}{r^s})$ among the perturbation around ${\stackrel{\circ}{\alpha}}{}^\mu$, into (\ref{B}) on shell to get the energy-momentum quasilocalization
\begin{equation}
-\alpha^\mu\Delta\omega^{\alpha\beta}\wedge\eta_{\alpha\beta\mu}-{\stackrel{\circ}{\textrm{D}}}{}^{\nu}{\stackrel{\circ}{\alpha}}{}^{\mu}\Delta\eta_{\mu\nu}=\alpha^\mu p_\mu,
\end{equation}
in which $p^0=-\Delta\omega^{\alpha\beta}\wedge\eta_{\alpha\beta}{}^0$ is the energy quasilocalization and $p^k=-\Delta\omega^{\alpha\beta}\wedge\eta_{\alpha\beta}{}^k$, where $k$ runs from $1$ to $3$, is the momentum quasilocalization in the $k$ direction.

For the angular momentum and center-of-mass moment, we prescribe the asymptotic displacement to be the asymptotic Poincar\'{e} rotation-boost $N^\mu=\varepsilon^\mu{}_\nu x^\nu\approx[{\stackrel{\circ}{\varepsilon}}{}^\mu{}_\nu+\varepsilon^\mu{}_\nu{}^+_1+\varepsilon^\mu{}_\nu{}^-_2]x^\nu$ when expanded in an asymptotic Lorentz frame, in which $\varepsilon_{\mu\nu}=\varepsilon_{[\mu\nu]}$ are the rotation-boost parameters with ${\stackrel{\circ}{\varepsilon}}{}^\mu{}_\nu$ being its values in Minkowski space and $x^\nu$ are components of the position vector, into (\ref{B}) on shell to get the angular momentum (center-of-mass moment) quasilocalization
\begin{eqnarray}\label{L}
&&-\varepsilon^\mu{}_\nu\Delta\omega^{\alpha\beta}\wedge x^\nu\eta_{\alpha\beta\mu}-{\stackrel{\circ}{\varepsilon}}{}^\mu{}_\nu\Delta\eta_\mu{}^\nu \nonumber \\
&&\sim-\frac{1}{2}\varepsilon^{\mu\nu}(x_\nu\Delta\omega^{\alpha\beta}\wedge\eta_{\alpha\beta\mu}-\Delta\omega^{\alpha\beta}\wedge\eta_{\alpha\beta\nu} x_\mu+\Delta\eta_{\mu\nu}-\Delta\eta_{\nu\mu}) \nonumber \\
&&=\frac{1}{2}\varepsilon^{\mu\nu}L_{\mu\nu}.
\end{eqnarray}
Among the above, we write ${\stackrel{\circ}{\varepsilon}}{}^\mu{}_\nu$ in the second term of the first line as $\varepsilon^\mu{}_\nu$ in the second line because $\Delta\varepsilon^\mu{}_\nu\Delta\eta_\mu{}^\nu$ makes no contribution when integrated asymptotically if we take the asymptotic falloffs with parities of the spatial components of the coframe $\vartheta^\mu$ expanded in a certain asymptotic Lorentz frame to be $\vartheta^\mu\approx{\stackrel{\circ}{\vartheta}}{}^\mu+{\vartheta^{\mu}}^+_1+{\vartheta^{\mu}}^-_2$, and consider that a variable of even parity, when integrated over a 2-surface bordering an asymptotically flat spatial region, makes no contribution because the components of the area 2-form $\textrm{i}_\textbf{\textit{n}}\rm{vol}^3$ expanded in a certain Cartesian frame are of odd parity. Here $\rm{vol}^3$ designates the spatial 3-volume form and $\textbf{\textit{n}}$ the unit normal to the boundary 2-surface.
In (\ref{L}), $L^{jk}=-x^k\Delta\omega^{\alpha\beta}\wedge\eta_{\alpha\beta}{}^j+\Delta\omega^{\alpha\beta}\wedge\eta_{\alpha\beta}{}^k x^j-\Delta\eta^{jk}+\Delta\eta^{kj}$ is the angular momentum quasilocalization in the $jk$ plane, and $L^{0k}=-x^k\Delta\omega^{\alpha\beta} \wedge \eta_{\alpha\beta}{}^0+\Delta\omega^{\alpha\beta} \wedge \eta_{\alpha\beta}{}^k x^0-\Delta\eta^{0k}+\Delta\eta^{k0}$ is the center-of-mass moment quasilocalization along the $k$ direction.

When $\mathcal{H}$ is integrated over an asymptotically flat spatial 3-hypersurface $\rm{S}^3$ with boundary, we gain the conservative quantity $H=\int_{\rm{S}}\mathcal{H}=\int_{\rm{S}}\textrm{i}_{\it{h}}\eta=\int_{\rm{S}}(h^0\vartheta^1\wedge\vartheta^2\wedge\vartheta^3-h^1\vartheta^0\wedge\vartheta^2\wedge \vartheta^3+h^2\vartheta^0\wedge \vartheta^1\wedge \vartheta^3-h^3\vartheta^0\wedge \vartheta^1\wedge \vartheta^2)=\int_{\rm{S}}h^0\vartheta^1\wedge\vartheta^2\wedge\vartheta^3=\int_{\rm{S}}h^0\rm{vol}^3$, where $h$ is a 4-vector. This can be seen as follows. Given the above prescription, we arrive at $\textrm{d}\mathcal{H}=\textrm{div}h\ \eta$. Because $\textrm{d}\mathcal{H}=0$ and $\eta\neq0$, we conclude $\textrm{div}h=\frac{\partial h^0}{\partial t}+\nabla\cdot\textbf{\textit{h}}=0$, where $\textbf{\textit{h}}$ is the spatial 3-vector of $h$. Assuming $\textbf{\textit{h}}$ vanishes in spatial infinity, we have $\frac{\textrm{d}H}{\textrm{d}t}=\frac{\textrm{d}}{\textrm{d}t}\int_{\rm{S}}h^0\rm{vol}^3=\int_{\rm{S}}\frac{\partial \textit{h}^0}{\partial \textit{t}}\rm{vol}^3=-\int_{\rm{S}}\nabla\cdot\textbf{\textit{h}}\textrm{vol}^3=-\int_{\partial S}\textbf{\textit{h}}\cdot\textbf{\textit{n}}\ \textrm{i}_\textbf{\textit{n}}\rm{vol}^3=0$. While the integrals of $p_\mu$ and $L_{\mu\nu}$ over an 2-surface bordering an asymptotically flat space give the conservative total quantities, their integrals over a finite closed orientable spatial 2-surface which is not too far from flat space give the corresponding quasilocal quantities. Note that the integral values only depend on the variable values taken on the boundary and are thus quasilocal.
In some angular momentum cases, ${\stackrel{\circ}{\textrm{D}}}{}^{\beta}{\stackrel{\circ}{N}}{}^{\alpha}\Delta\eta_{\alpha\beta}$ is not added, such as the theme of this work, the spinor Hamiltonians we will discuss later.

To get the field equations and boundary condition, we vary (\ref{HF}) along with (\ref{B}):
\begin{eqnarray}\label{varyHamiltonian}
&&\delta\mathcal{H}_\vartheta=R^{\alpha\beta}\wedge\delta \textrm{i}_N\eta_{\alpha\beta}-\textrm{D}\textrm{i}_N\omega^{\alpha\beta}\wedge\delta\eta_{\alpha\beta} \nonumber \\
&&\qquad\>\>\>\ +\delta \textrm{i}_N\omega^{\alpha\beta} \textrm{D}\eta_{\alpha\beta}-\delta\omega^{\alpha\beta}\wedge(- \textrm{D}\textrm{i}_N\eta_{\alpha\beta}+\textrm{i}_N\omega^\gamma{}_\alpha\eta_{\beta\gamma}-\textrm{i}_N\omega^\gamma{}_\beta\eta_{\alpha\gamma}) \nonumber \\
&&\qquad\>\>\>\ +\textrm{d}(-\Delta\omega^{\alpha\beta}\wedge\delta \textrm{i}_N\eta_{\alpha\beta}+\textrm{i}_N\Delta\omega^{\alpha\beta}\delta\eta_{\alpha\beta}).
\end{eqnarray}
When the integral of the boundary term, the superpotential of the third line of (\ref{varyHamiltonian}), vanishes, the first and second lines give the field equations.
On the other hand, we see that the Hamiltonian variation boundary term has the symplectic structure, which, according to the boundary variation principle, reveals which variables are to be controlled and which variables are the responses. In this case, we should control (certain projected components of) the coframe, which is geometrically equivalent to controlling the metric, and leave (certain projected components of) the connection as the response. The replacement of the Hamiltonian boundary term $\mathcal{B}_\vartheta$ has changed the boundary condition; the original boundary condition is, as shown in the boundary term of $\delta\mathcal{L}_{\rm{EC}}$, is that the connection is the control variable and the coframe is the response variable.

The theme of this paper is to explore the Witten spinor Hamiltonian adapted for the angular momentum. We want to find the spinor Hamiltonian 3-form which can give the angular momentum quasilocalization for a gravitational field. The properties that we expect a spinor Hamiltonian to satisfy are: for an asymptotically flat spatial 3-region, (i) the Hamiltonian generates correct dynamical evolution equations; if the variables have appropriately asymptotic falloffs with parities, (ii) the part of the boundary integral of the Hamiltonian variation involved with the variation of the dynamical variables vanishes, (iii) the Hamiltonian boundary integral gives finite but nonvanishing values, and (iv) the Hamiltonian boundary integral gives desired values for the energy-momentum, angular momentum and center-of-mass moment when the displacement vector $N$ is suitably chosen; these mean that the Hamiltonian 3-form has the pattern given by (\ref{HF}) with the boundary term asymptotically approaching (\ref{B}). We will try to compose a spinor Hamiltonian for the angular momentum according to these criteria.

\section{The Dirac algebra, Dirac spinor and Clifforms}\label{Spinor}

In the following sections, we will use the Dirac spinor and Clifforms to formulate our Hamiltonians, so let us give a brief overview of them.

Introducing a pseudo-orthonormal coframe field $\vartheta^\mu$ endued with a spin structure to an asymptotically flat spacetime region $\textrm{M}^4$ \footnote{For our spinor formulation, we assume $\textrm{M}^4$ admits spin structures. The conditions necessary and sufficient for a pseudo-Riemannian manifold to admit spin structures are orientability and parallelizability.}, we can associate $\vartheta^\mu$ with a Dirac algebra $\textrm{C}_4$ that is generated by the Dirac matrices $\gamma_\lambda$ satisfying
\begin{equation}\label{Clifford}
\gamma_\alpha\gamma_\beta+\gamma_\beta\gamma_\alpha=2g_{\alpha\beta}\rm{I},
\end{equation}
where $g_{\alpha\beta}$ are the metric components in $\vartheta^\mu$. We can raise the index of a Dirac matrix by $\gamma^\alpha=g^{\alpha\beta}\gamma_\beta$, $g^{\alpha\beta}$ being the components of the inverse of the metric. Then given the notation $\gamma^{\alpha\beta...}:=\gamma^{[\alpha\beta...]}$, the volume form of the Dirac algebra is defined to be
\begin{equation}
\gamma:=\gamma^{0123}=\gamma^0\gamma^1\gamma^2\gamma^3=-\gamma_0\gamma_1\gamma_2\gamma_3,
\end{equation}
the second $``="$ being able to be explained by taking $\alpha\neq\beta$ in (\ref{Clifford}).
And it is easy to calculate that $\gamma\gamma=-1$. Identities which will be employed are
\begin{eqnarray}
&&\gamma_\lambda\gamma_{\mu\nu}=\gamma_{\lambda\mu\nu}+g_{\lambda\mu}\gamma_\nu-g_{\lambda\nu}\gamma_\mu, \label{DiracAl1} \\
&&\gamma_{\mu\nu}\gamma_\lambda=\gamma_{\mu\nu\lambda}-g_{\mu\lambda}\gamma_\nu+g_{\nu\lambda}\gamma_\mu, \label{DiracAl2} \\
\textrm{and} \ \ &&\gamma_\lambda\gamma_{\mu\nu}+\gamma_{\mu\nu}\gamma_\lambda=2\gamma_{\lambda\mu\nu}=-2\eta_{\lambda\mu\nu\kappa}\gamma\gamma^\kappa. \label{DiracAl3}
\end{eqnarray}
The first $``="$ in (\ref{DiracAl3}) comes from the addition of (\ref{DiracAl1}) and (\ref{DiracAl2}) while the second $``="$ is explained as follows: We can write $\gamma=-\gamma_{0123}$. The indices $\lambda\mu\nu$ of $\eta$ determine the sign resulting from the order change from the increasing to $\lambda\mu\nu$ and $\kappa$ picks the index which is not included in $\lambda\mu\nu$ among $0123$ to make $\gamma_\kappa\gamma^\kappa=1$, where no summation over $\kappa$ is implied.

The 2-grade subalgebra of $\rm{C}_4$ spans the Lie algebra of a spinor representation for the proper orthochronous Lorentz group $\rm{L}_0$, namely the direct sum of two inequivalent $\rm{SL}(2, \mathbf{C})$ representations for $\textrm{L}_0$, $A$ and $A^{\dagger^-1}$. This representation is labeled as $\mathcal{D}(\frac{1}{2}, \frac{1}{2})$ when expressed in terms of $\rm{SU}(2, \mathbf{C})\times \rm{SU}(2, \mathbf{C})$\cite{Tung} and its associated bundle is the Dirac spinor bundle. Given a Dirac spinor $\psi$, we define the spinor adjoint to it as $\overline{\psi}:=\psi^\dagger\beta$, where $\beta$ defines a metric for the Dirac spinor bundle, being a matrix in $\rm{C}_4$ with which two conditions are satisfied: first, $\overline{\psi}\psi$, parameterizing a Lorentz scalar, is real, that is tantamount to $\beta^\dagger=\beta$; second, $\overline{\psi}\gamma^\mu\psi$, parameterizing the components of a Lorentz vector, are real, that is parallel to $(\beta\gamma^\mu)^\dagger=\beta\gamma^\mu$. It can be shown that $\gamma_0$ satisfies both conditions and can be a choice for $\beta$.

The Dirac spinor together with the Dirac matrices can be used to parameterize Lorentz (pseudo-)tensors \cite{Ryder}. In addition to the aforementioned properties that $\overline{\psi}\psi$ parameterizes a scalar and $\overline{\psi}\gamma^\mu\psi$ parameterizes the $\mu$ component of a vector, the latter characterizing a Poincar\'{e} translation displacement, there are still the followings.
\begin{itemize}
  \item $\textrm{i}\overline{\psi}\gamma^\mu\gamma\psi$ parameterizes a Lorentz pseudovector, i.e. it behaves the same as a Lorentz vector except that under space inversion (parity) its temporal component changes the sign while its spatial components don't.
  \item $\textrm{i}\overline{\psi}\gamma_{\mu\nu}\psi$ parameterizes a Lorentz bivector, which characterizes a rotation-boost parameter.
  \item $\textrm{i}\overline{\psi}\gamma_{\mu\nu}\gamma\psi$ parameterizes a Lorentz pseudo-bivector.
\end{itemize}

The above Clifford algebra $\rm{C}_4$ can be generalized to the Clifford-algebra-valued $r$-forms with $r$ being a non-negative integer, abbreviated as Clifforms \cite{Clifform}. A general Clifform can be written as $\textsf{T}=T^{\alpha\beta ...}\gamma_{\alpha\beta ...}$ with $T^{\alpha\beta ...}$ being the $r$-form coefficient, defined on the vectors of the tangent bundle to $\textrm{M}^4$, of $\gamma_{\alpha\beta ...}$. When $r=0$, Clifforms reduce to the Clifford algebra. One Clifform of our concern is the connection $\omega:=\frac{1}{4}\omega^{\alpha\beta}\gamma_{\alpha\beta}$ for the Dirac spinor bundle. So the first covariant differentials of $\psi$ and $\overline{\psi}$ are
\begin{eqnarray}
&&\textrm{D}\psi=\textrm{d}\psi+\omega\wedge\psi \\
\rm{and} \ \ &&\textrm{D}\overline{\psi}=\textrm{d}\overline{\psi}-\overline{\psi}\wedge\omega,
\end{eqnarray}
respectively.
The second covariant differentials of $\psi$ and $\overline{\psi}$ are
\begin{eqnarray}
&&\textrm{D}^2\psi=\textsf{R}\wedge\psi\label{2CDS} \\
\rm{and} \ \ &&\textrm{D}^2\overline{\psi}=-\overline{\psi}\wedge \textsf{R},\label{2CDJS}
\end{eqnarray}
respectively, where $\textsf{R}:=\frac{1}{4}R^{\alpha\beta}\gamma_{\alpha\beta}=\textrm{d}\omega+\omega\wedge\omega\equiv \textrm{d}\omega+\frac{1}{2}[\omega, \omega]$ is the curvature for the Dirac spinor bundle, another Clifform of our concern.
Still another Clifform which is of our concern is $\vartheta:=\vartheta^\alpha\gamma_\alpha$.

Finally in this section we come to the generic form of the key identity which underlies our spinor Hamiltonian formulation for transforming the quadratic spinor terms into the curvature term and total differential, the spinor-curvature identity \cite{SCI}: if $\phi$ and $\psi$ are two spinors while $\textsf{A}$ and $\textsf{B}$ are two Clifforms, in which the degree of the differential form part of $\textsf{A}$ is $a$, we have
\begin{eqnarray}\label{SCI}
&&\textrm{d}[\overline{\phi}\textsf{A}\wedge \textrm{D}(\textsf{B}\psi)-(-1)^a \textrm{D}(\overline{\phi}\textsf{A})\wedge \textsf{B}\psi]\nonumber \\
&&=2\textrm{D}(\overline{\phi}\textsf{A})\wedge \textrm{D}(\textsf{B}\psi)+2(-1)^a\overline{\phi}\textsf{A}\wedge \textsf{R}\wedge \textsf{B}\psi,
\end{eqnarray}
which results from $\textrm{D}\alpha=\textrm{d}\alpha$ when $\alpha$ is a scalar, the employment of (\ref{2CDS}) and (\ref{2CDJS}) as well as
$\textrm{d}\gamma_{\alpha\beta...}=0$ and $\textrm{D}\gamma_{\alpha\beta...}=0$ \footnote{This identity holds true under the condition of metric compatibility besides the adoption of (a)an (pseudo-)orthonormal frame. Our spinor Hamiltonian formulation for GR is in such a context.}.

\section{Failure of the pseudovectorial parameterization in Witten spinor Hamiltonian for angular momentum}\label{Pseudovector}

Witten and Nester conceived the quadratic spinor Hamiltonian 3-form  \cite{Witten,PositiveEnergy,SpinorFormulation,Straumann},
\begin{eqnarray}\label{WittenHamiltonian}
&&\mathcal{H}_{\rm{w}}(\psi):=2[\textrm{D}\overline{\psi}\wedge \gamma \textrm{D}(\vartheta\psi)+\textrm{D}(\overline{\psi}\vartheta)\gamma\wedge \textrm{D}\psi] \nonumber \\
&&\qquad\quad\>\;\equiv N^\mu R^{\alpha\beta}\wedge\eta_{\alpha\beta\mu}\nonumber \\
&&\qquad\qquad\>\;+\textrm{d}[(\overline{\psi}\vartheta\gamma)\wedge \textrm{D}\psi-\textrm{D}\overline{\psi}\wedge(\gamma\vartheta\psi)+\textrm{D}(\overline{\psi}\vartheta)\gamma\psi +\overline{\psi}\gamma \textrm{D}(\vartheta\psi)], \nonumber \\
\end{eqnarray}
with the spinor vectorial parameterization $N^\mu=\overline{\psi}\gamma^\mu\psi$ for the translation displacement. Its boundary term yields a gravitational energy-momentum quasilocalization. When Einstein's vacuum field equations are satisfied, constraining the spinor to satisfy the Witten equation, $\mathcal{H}_{\rm{w}}$ gives a positive energy proof. The construction of $\mathcal{H}_{\rm{w}}$ relies on the spinor-curvature identity, which originated from the idea of the Sparling 3-form \cite{Sparling1982,SparlingForm,SparlingAshtekar,SparlingSzabados,PenroseRindler}, $S_\mu=-2G^\nu{}_\mu\eta_\nu-dW_\mu$, where $G^\nu{}_\mu=\texttt{R}^\nu{}_\mu-\frac{1}{2}g^\nu{}_\mu R$ with $\texttt{R}_{\nu\mu}$ being the Ricci curvature and $R$ being the (Ricci) scalar curvature, for $R^{\alpha\beta}\wedge\eta_{\alpha\beta\mu}=-2G^\nu{}_\mu\eta_\nu$. Remarkably, $W_\mu$ is the antisymmetric part of the Pauli-Lubanski spin tensor, whose symmetric part gives the Fefferman conformal structure of the twistor Cauchy-Riemannian structure of the associated spacelike hypersurface. We want to see if an expression for the quasilocal angular momentum can also be constructed in this way.

For the angular momentum, we need a displacement which characterizes a rotation. Therefore we consider the Hamiltonian
\begin{eqnarray}
&&\mathcal{H}_{\rm{wa}}(\psi):=2[\textrm{i}\textrm{D}\overline{\psi}\wedge \textrm{D}(\vartheta\psi)-\textrm{i}\textrm{D}(\overline{\psi}\vartheta)\wedge \textrm{D}\psi] \nonumber \\
&&\qquad\quad\>\>\>\;\equiv N^\mu R^{\alpha\beta}\wedge\eta_{\alpha\beta\mu}\nonumber \\
&&\qquad\qquad\>\>\>\;+\textrm{d}[-\textrm{i}(\overline{\psi}\vartheta)\wedge \textrm{D}\psi-\textrm{i}\textrm{D}\overline{\psi}\wedge(\vartheta\psi)-\textrm{i}\textrm{D}(\overline{\psi}\vartheta)\psi+\textrm{i}\overline{\psi} \textrm{D}(\vartheta\psi)]
\end{eqnarray}
with the spinor pseudovectorial parameterization $N^\mu=\textrm{i}\overline{\psi}\gamma^\mu\gamma\psi$, which should be equal to $\varepsilon^\mu{}_\nu x^\nu$. But even though the angular momentum $L^i:=1/2\epsilon^{ijk}L_{jk}$, where $\epsilon^{ijk}$ is the Levi-Civita symbol, is a pseudovector, the rotational displacement $\varepsilon^\mu{}_\nu x^\nu$ is a vector, not a pseudovector! It is a contradictory parameterization. Thus this scheme of the spinor parameterization for the angular momentum is certainly infeasible.

What if we express the rotational displacement in terms of the spinor vectorial parameterization, that is, $\overline{\psi}\gamma^\mu\psi=\varepsilon^\mu{}_\nu x^\nu$? Let's take a look at it. This parameterization entails $\psi\approx\overline{\psi}\approx\sqrt{r}$, which would make the Hamiltonian 3-form (\ref{WittenHamiltonian}) have a divergent integral. So this parameterization for the rotational displacement is infeasible, either.

Since we can't succeed in considering these simple trials of adjusting the Witten Hamiltonian for the angular momentum, we consider another conceived quadratic spinor Hamiltonian, which has a satisfactory spinor parameterization for the rotation displacement, in next section.

\section{Witten spinor Hamiltonian adapted for angular momentum with the bivectorial parameterization, full 4-term}\label{4-term}

By the application of the spinor-curvature identity (\ref{SCI}), we conceive the following identity:
\begin{eqnarray}
&&\qquad\textrm{d}[\overline{\psi}\textrm{D}(\vartheta \textrm{i}\gamma{\not\!\! X}\psi)-\textrm{D}(\overline{\psi}{\not\!\! X}\textrm{i}\gamma\vartheta)\psi+(\overline{\psi}{\not\!\! X}\textrm{i}\gamma) \textrm{D}(\vartheta\psi)-\textrm{D}(\overline{\psi}\vartheta) (\textrm{i}\gamma {\not\!\! X}\psi) \nonumber \\
&&\qquad\ \ \>-(\overline{\psi}\vartheta)\wedge \textrm{D}(\textrm{i}\gamma{\not\!\! X}\psi)-\textrm{D}(\overline{\psi}{\not\!\! X}\textrm{i}\gamma)\wedge(\vartheta\psi)-(\overline{\psi}{\not\!\! X}\textrm{i}\gamma\vartheta)\wedge \textrm{D}\psi-\textrm{D}\overline{\psi}\wedge (\vartheta \textrm{i}\gamma {\not\!\! X}\psi)] \nonumber \\
&&\qquad\equiv2[\overline{\psi}\textsf{R}\wedge(\vartheta \textrm{i}\gamma{\not\!\! X}\psi)+(\overline{\psi}{\not\!\! X}\textrm{i}\gamma\vartheta)\wedge \textsf{R}\psi \nonumber \\
&&\qquad\quad\quad+(\overline{\psi}{\not\!\! X}\textrm{i}\gamma) \textsf{R}\wedge(\vartheta\psi)+(\overline{\psi}\vartheta)\wedge \textsf{R}(\textrm{i}\gamma {\not\!\! X}\psi)] \nonumber \\
&&\qquad\quad+2[\textrm{D}\overline{\psi}\wedge \textrm{D}(\vartheta \textrm{i}\gamma {\not\!\! X}\psi)-\textrm{D}(\overline{\psi}{\not\!\! X} \textrm{i}\gamma\vartheta)\wedge \textrm{D}\psi \nonumber \\
&&\qquad\qquad\ \ +\textrm{D}(\overline{\psi}{\not\!\! X}\textrm{i}\gamma)\wedge \textrm{D}(\vartheta\psi)-\textrm{D}(\overline{\psi}\vartheta)\wedge \textrm{D}(\textrm{i}\gamma{\not\!\! X}\psi)] \nonumber \\
&&\qquad=2\overline{\psi}[\textsf{R}\wedge\vartheta+\vartheta\wedge \textsf{R}) \textrm{i}\gamma {\not\!\! X}+{\not\!\! X} \textrm{i}\gamma(\vartheta\wedge \textsf{R}+\textsf{R}\wedge\vartheta)]\psi \nonumber \\
&&\qquad\quad+2[\textrm{D}\overline{\psi}\wedge \textrm{D}(\vartheta \textrm{i}\gamma {\not\!\! X}\psi)-\textrm{D}(\overline{\psi}{\not\!\! X} \textrm{i}\gamma\vartheta)\wedge \textrm{D}\psi\nonumber \\
&&\qquad\qquad\ \ +\textrm{D}(\overline{\psi}{\not\!\! X}\textrm{i}\gamma)\wedge \textrm{D}(\vartheta\psi)-\textrm{D}(\overline{\psi}\vartheta)\wedge \textrm{D}(\textrm{i}\gamma{\not\!\! X}\psi)],
\end{eqnarray}
in which ${\not\!\! X}:=x^\nu\gamma_\nu$.
Among the above,
\begin{eqnarray}\label{repetitivepattern}
&&\overline{\psi}[\textsf{R}\wedge\vartheta+\vartheta\wedge \textsf{R}) \textrm{i}\gamma {\not\!\! X}+{\not\!\! X} \textrm{i}\gamma(\vartheta\wedge \textsf{R}+\textsf{R}\wedge\vartheta)]\psi \nonumber \\
&&=\frac{1}{4}R^{\alpha\beta}\wedge\vartheta^\lambda\overline{\psi}[(\gamma_{\alpha\beta}\gamma_\lambda+\gamma_\lambda\gamma_{\alpha\beta})\textrm{i}\gamma {\not\!\! X}+{\not\!\! X}\textrm{i}\gamma(\gamma_\lambda\gamma_{\alpha\beta}+\gamma_{\alpha\beta}\gamma_\lambda)]\psi \nonumber \\
&&=\frac{\textrm{i}}{4}R^{\alpha\beta}\wedge\vartheta^\lambda\overline{\psi}[2\gamma_{\lambda\alpha\beta}\gamma{\not\!\! X}+{\not\!\! X}\gamma 2\gamma_{\lambda\alpha\beta}]\psi \nonumber \\
&&=\frac{\textrm{i}}{2}R^{\alpha\beta}\wedge\vartheta^\lambda\overline{\psi}[ -\eta_{\lambda\alpha\beta\kappa}\gamma\gamma^\kappa\gamma{\not\!\! X}-{\not\!\! X}\gamma\eta_{\lambda\alpha\beta\kappa}\gamma\gamma^\kappa]\psi \nonumber \\
&&=R^{\alpha\beta}\wedge\eta_{\alpha\beta\kappa}\frac{\textrm{i}}{2}\overline{\psi}[\gamma^\kappa{\not\!\! X}-{\not\!\! X}\gamma^\kappa]\psi.
\end{eqnarray}
Here we reach the result,
\begin{eqnarray}\label{repetitivepattern}
\quad\ \vartheta^\lambda\overline{\psi}[(\gamma_{\alpha\beta}\gamma_\lambda+\gamma_\lambda\gamma_{\alpha\beta})\textrm{i}\gamma {\not\!\! X}+{\not\!\! X}\textrm{i}\gamma(\gamma_\lambda\gamma_{\alpha\beta}+\gamma_{\alpha\beta}\gamma_\lambda)]\psi=2\textrm{i}\eta_{\alpha\beta\kappa}\overline{\psi}[\gamma^\kappa{\not\!\! X}-{\not\!\! X}\gamma^\kappa]\psi, \nonumber \\
\end{eqnarray}
which will be used again later.
Thus we can compose the spinor Hamiltonian 3-form for the angular momentum,
\begin{eqnarray}\label{HA4}
&&\mathcal{H}(\psi, {\not\!\! X}):=-\textrm{D}\overline{\psi}\wedge \textrm{D}(\vartheta \textrm{i}\gamma {\not\!\! X}\psi)+\textrm{D}(\overline{\psi}{\not\!\! X} \textrm{i}\gamma\vartheta)\wedge \textrm{D}\psi \nonumber \\
&&\qquad\qquad\quad\,-\textrm{D}(\overline{\psi}{\not\!\! X}\textrm{i}\gamma)\wedge \textrm{D}(\vartheta\psi)+\textrm{D}(\overline{\psi}\vartheta)\wedge \textrm{D}(\textrm{i}\gamma{\not\!\! X}\psi) \nonumber \\
&&\qquad\qquad\>\equiv R^{\alpha\beta}\wedge\eta_{\alpha\beta\kappa}\frac{\textrm{i}}{2}\overline{\psi}[\gamma^\kappa{\not\!\! X}-{\not\!\! X}\gamma^\kappa]\psi \nonumber \\
&&\qquad\qquad\quad\>+\frac{1}{2}\textrm{d}[-\overline{\psi}\textrm{D}(\vartheta \textrm{i}\gamma{\not\!\! X}\psi)+\textrm{D}(\overline{\psi}{\not\!\! X}\textrm{i}\gamma\vartheta)\psi \nonumber \\
&&\qquad\qquad\quad\>\quad\quad\ \,-(\overline{\psi}{\not\!\! X}\textrm{i}\gamma) \textrm{D}(\vartheta\psi)+\textrm{D}(\overline{\psi}\vartheta) (\textrm{i}\gamma {\not\!\! X}\psi) \nonumber \\
&&\qquad\qquad\quad\>\quad\quad\ \,+(\overline{\psi}\vartheta)\wedge \textrm{D}(\textrm{i}\gamma{\not\!\! X}\psi)+\textrm{D}(\overline{\psi}{\not\!\! X}\textrm{i}\gamma)\wedge(\vartheta\psi) \nonumber \\
&&\qquad\qquad\quad\>\quad\quad\ \,+(\overline{\psi}{\not\!\! X}\textrm{i}\gamma\vartheta)\wedge \textrm{D}\psi+\textrm{D}\overline{\psi}\wedge (\vartheta \textrm{i}\gamma {\not\!\! X}\psi)]
\end{eqnarray}
with
\begin{eqnarray}
N^\kappa=\varepsilon^\kappa{}_\nu x^\nu=\overline{\psi}\textrm{i}\gamma^\kappa{}_\nu\psi x^\nu=\frac{\textrm{i}}{2}\overline{\psi}[\gamma^\kappa{\not\!\! X}-{\not\!\! X}\gamma^\kappa]\psi,
\end{eqnarray}
for a Poincar\'{e} rotational displacement, in which we express the rotational parameters in terms of the spinor bivectorial parameterization: $\varepsilon^\kappa{}_\nu=\overline{\psi}\textrm{i}\gamma^\kappa{}_\nu\psi$.

Actually the boundary 2-form of this Hamiltonian 3-form has an infinite number of expressions. Let us pause to take a look at it:
\begin{eqnarray}\label{I1}
\textrm{dd}[\overline{\psi}(\vartheta \textrm{i}\gamma{\not\!\! X}\psi)]=0 &\Longrightarrow& \textrm{d}[\textrm{D}\overline{\psi}\wedge(\vartheta \textrm{i}\gamma{\not\!\! X}\psi)+\overline{\psi}\textrm{D}(\vartheta \textrm{i}\gamma{\not\!\! X}\psi)]=0 \nonumber \\
&\Longrightarrow& \textrm{d}[\textrm{D}\overline{\psi}\wedge(\vartheta \textrm{i}\gamma{\not\!\! X}\psi)]=-\textrm{d}[\overline{\psi}\textrm{D}(\vartheta \textrm{i}\gamma{\not\!\! X}\psi)].
\end{eqnarray}
By the same principle, we can also get
\begin{eqnarray}
&&\textrm{d}[\textrm{D}(\overline{\psi}{\not\!\! X}\textrm{i}\gamma\vartheta)\psi]=\textrm{d}[(\overline{\psi}{\not\!\! X}\textrm{i}\gamma\vartheta)\wedge \textrm{D}\psi], \label{I2}\\
&&\textrm{d}[\textrm{D}(\overline{\psi}{\not\!\! X}\textrm{i}\gamma)\wedge(\vartheta\psi)]=-\textrm{d}[(\overline{\psi}{\not\!\! X}\textrm{i}\gamma) \textrm{D}(\vartheta\psi)] \label{I3} \\
\textrm{and}\ &&\textrm{d}[\textrm{D}(\overline{\psi}\vartheta) (\textrm{i}\gamma {\not\!\! X}\psi)]=\textrm{d}[(\overline{\psi}\vartheta)\wedge \textrm{D}(\textrm{i}\gamma{\not\!\! X}\psi)]. \label{I4}
\end{eqnarray}
Additionally, the Hamiltonian boundary expression is expected to be real to adapt to the real requirement of physical quantities, so we need to restrict it to be the linear combination of
\begin{eqnarray}
&&-\overline{\psi}\textrm{D}(\vartheta \textrm{i}\gamma{\not\!\! X}\psi)+\textrm{D}(\overline{\psi}{\not\!\! X}\textrm{i}\gamma\vartheta)\psi, \label{set1}\\
&&-(\overline{\psi}{\not\!\! X}\textrm{i}\gamma) \textrm{D}(\vartheta\psi)+\textrm{D}(\overline{\psi}\vartheta) (\textrm{i}\gamma {\not\!\! X}\psi), \label{set2}\\
&&(\overline{\psi}\vartheta)\wedge\textrm{D}(\textrm{i}\gamma{\not\!\! X}\psi)+\textrm{D}(\overline{\psi}{\not\!\! X}\textrm{i}\gamma)\wedge(\vartheta\psi)\label{set3} \\
\textrm{and} \qquad
&&(\overline{\psi}{\not\!\! X}\textrm{i}\gamma\vartheta)\wedge \textrm{D}\psi+\textrm{D}\overline{\psi}\wedge(\vartheta \textrm{i}\gamma {\not\!\! X}\psi), \label{set4}
\end{eqnarray}
in which each pair consists of two terms complex conjugate to each other \footnote{Though $\overline{\psi}\textrm{D}({\not\!\! X}\textrm{i}\gamma \vartheta\psi)-\textrm{D}(\overline{\psi}\vartheta\textrm{i}\gamma{\not\!\! X})\psi$ and $(\overline{\psi}\vartheta\textrm{i}\gamma{\not\!\! X})\wedge\textrm{D}\psi+\textrm{D}\overline{\psi}\wedge({\not\!\! X}\textrm{i}\gamma \vartheta\psi)$ are two possible pairs complex conjugate to each other, they are not employed, for they don't lead to desired terms after rearrangements.}. The coefficients of the four pairs need to be manipulated so that the combination approaches the ADM term asymptotically. Once a valid expression is reached, we can use (\ref{I1}), (\ref{I2}), (\ref{I3}) and (\ref{I4}) to attain to the alternative Hamiltonian boundary expressions: we can gain the alternative expressions for (\ref{set1}) by adding a constant multiple of
\begin{eqnarray}\label{adding1}
&&[\textrm{D}\overline{\psi}\wedge (\vartheta \textrm{i}\gamma {\not\!\! X}\psi)+\overline{\psi} \textrm{D}(\vartheta \textrm{i}\gamma {\not\!\! X}\psi)-\textrm{D}(\overline{\psi}{\not\!\! X}\textrm{i}\gamma\vartheta)\psi+(\overline{\psi}{\not\!\! X}\textrm{i}\gamma\vartheta)\wedge \textrm{D}\psi] \nonumber \\
&&=\textrm{d}[\overline{\psi}(\vartheta \textrm{i}\gamma{\not\!\! X}-{\not\!\! X}\textrm{i}\gamma\vartheta)\psi],
\end{eqnarray}
the alternative expressions for (\ref{set2}) by adding a constant multiple of
\begin{eqnarray}\label{adding2}
&&\qquad[-\textrm{D}(\overline{\psi}\vartheta) (\textrm{i}\gamma {\not\!\! X}\psi)+(\overline{\psi}\vartheta)\wedge\textrm{D}(\textrm{i}\gamma{\not\!\! X}\psi)+\textrm{D}(\overline{\psi}{\not\!\! X}\textrm{i}\gamma)\wedge(\vartheta\psi)+(\overline{\psi}{\not\!\! X}\textrm{i}\gamma) \textrm{D}(\vartheta\psi)] \nonumber \\
&&\qquad=-\textrm{d}[\overline{\psi}(\vartheta \textrm{i}\gamma{\not\!\! X}-{\not\!\! X}\textrm{i}\gamma\vartheta)\psi],
\end{eqnarray}
the alternative expressions for (\ref{set3}) by adding a constant multiple of
\begin{eqnarray}\label{adding3}
&&\qquad[\textrm{D}(\overline{\psi}\vartheta) (\textrm{i}\gamma {\not\!\! X}\psi)-(\overline{\psi}\vartheta )\wedge\textrm{D}(\textrm{i}\gamma{\not\!\! X}\psi)-\textrm{D}(\overline{\psi}{\not\!\! X}\textrm{i}\gamma)\wedge(\vartheta\psi)-(\overline{\psi}{\not\!\! X}\textrm{i}\gamma) \textrm{D}(\vartheta\psi)], \nonumber \\
&&\qquad=\textrm{d}[\overline{\psi}(\vartheta \textrm{i}\gamma{\not\!\! X}-{\not\!\! X}\textrm{i}\gamma\vartheta)\psi],
\end{eqnarray}
and the alternative expressions for (\ref{set4}) by adding a constant multiple of
\begin{eqnarray}\label{adding4}
&&\qquad[-\textrm{D}\overline{\psi}\wedge(\vartheta \textrm{i}\gamma {\not\!\! X}\psi)-\overline{\psi}\textrm{D}(\vartheta \textrm{i}\gamma{\not\!\! X}\psi)+\textrm{D}(\overline{\psi}{\not\!\! X}\textrm{i}\gamma\vartheta)\psi-(\overline{\psi}{\not\!\! X}\textrm{i}\gamma\vartheta)\wedge \textrm{D}\psi]\nonumber \\
&&\qquad=-\textrm{d}[\overline{\psi}(\vartheta \textrm{i}\gamma{\not\!\! X}-{\not\!\! X}\textrm{i}\gamma\vartheta)\psi],
\end{eqnarray}
each of which is a multiple of the same total differential. Therefore we can add an arbitrary real constant multiple (to make the addition ostensibly real) of $\textrm{d}[\overline{\psi}(\vartheta \textrm{i}\gamma{\not\!\! X}-{\not\!\! X}\textrm{i}\gamma\vartheta)\psi]$ to a valid expression to arrive at the alternatives \footnote{In fact, in addition to the multiple of this total differential, the boundary expression is still indefinite up to an arbitrary total differential because such an addition doesn't change the boundary integral nor the Hamiltonian 3-form.}.

This Hamiltonian 3-form (\ref{HA4}) has the desired volume term with the displacement characteristic of a rotation when the torsion is assumed to vanish \textit{a priori}. Then now let us check if it satisfies the other 3 principles, (ii), (iii) and (iv), addressed in the final paragraph of Sec. \ref{HamiltonianFormulation}.

First let us see if its boundary integral gives a finite value. We need to perturb the components of the dynamical variable $\vartheta$ and the spinor $\psi$ around their flat-space constant values (for the spinor, accurately speaking, it is the perturbation around the flat-space value of the displacement that in turn requires the perturbation around the constant spinor). Let us try to take the asymptotic falloffs with parities of the spatial components of the dynamical variables expanded in a certain asymptotic Lorentz frame as
\begin{eqnarray}
&&\vartheta\approx\stackrel{\circ}{\vartheta}+\vartheta^+_1+\vartheta^-_2 \\
&&\omega\approx d\vartheta\approx\omega^-_2+\omega^+_3,
\end{eqnarray}
and the asymptotic falloffs with parities of the components of the Dirac spinor expanded in a certain $\mathcal{D}(\frac{1}{2},\frac{1}{2})$-frame in the infinity as
\begin{eqnarray}
\overline{\psi}\approx\psi\approx\stackrel{\circ}{\psi}+\psi^+_1+\psi^-_2.
\end{eqnarray}
Let us check the asymptotic falloffs with parities of each term in the boundary 2-form of the Hamiltonian 3-form (\ref{HA4}) when the asymptotic falloffs with parities of fields are taken as above.
\begin{eqnarray}\label{fot1}
&&\quad\qquad\textrm{D}(\overline{\psi}{\not\!\! X}\textrm{i}\gamma\vartheta)\psi\approx\overline{\psi}\textrm{D}(\vartheta \textrm{i}\gamma{\not\!\! X}\psi)=\overline{\psi}\textrm{d}(\vartheta \textrm{i}\gamma{\not\!\! X}\psi)+\overline{\psi}\omega\wedge(\vartheta \textrm{i}\gamma{\not\!\! X}\psi) \nonumber \\
&&\quad\qquad \approx\quad[\stackrel{\circ}{\overline{\psi}}+\overline{\psi}^+_1+\overline{\psi}^-_2]\{{\not\!\! X}[\textrm{d}\vartheta^+_1+\textrm{d}\vartheta^-_2+\textrm{d}\psi^
+_1+\textrm{d}\psi^-_2+\textrm{d}(\vartheta^+_1\psi^
+_1)^+_2+...]\nonumber \\
&&\qquad\qquad\qquad\qquad\qquad\quad\ \ +\textrm{d}{\not\!\! X}[\stackrel{\circ}{\vartheta}+\vartheta^+_1+\vartheta^-_2][\stackrel{\circ}{\psi}+\psi^
+_1+\psi^-_2]\} \nonumber \\
&&\qquad\quad\ \ \ \ +[\stackrel{\circ}{\overline{\psi}}+\overline{\psi}^+_1+\overline{\psi}^-_2][\omega^-_2+\omega^+_3] [\stackrel{\circ}{\vartheta}+\vartheta^+_1+\vartheta^-_2]{\not\!\! X}[\stackrel{\circ}{\psi}+\psi^
+_1+\psi^-_2] \nonumber \\
&&\quad\qquad \approx{\not\!\! X}(\textrm{d}\vartheta^-_2+\textrm{d}\psi^-_2+\omega^+_3+...)+\textrm{d}{\not\!\! X}(\stackrel{\circ}{\overline{\psi}}\stackrel{\circ}{\vartheta}\stackrel{\circ}{\psi}+\overline{\psi}^-_2+\vartheta^-_2+\psi^-_2+...)
\end{eqnarray}
\hrulefill
\begin{eqnarray}\label{fot2}
&&\quad\qquad\textrm{D}(\overline{\psi}\vartheta) (\textrm{i}\gamma {\not\!\! X}\psi)\approx
(\overline{\psi}{\not\!\! X}\textrm{i}\gamma) \textrm{D}(\vartheta\psi)\nonumber \\
&&\quad\qquad \approx[\stackrel{\circ}{\overline{\psi}}+\overline{\psi}^+_1+\overline{\psi}^-_2]{\not\!\! X}\{[\textrm{d}\vartheta^+_1+\textrm{d}\vartheta^-_2+\textrm{d}\psi^
+_1+\textrm{d}\psi^-_2+\textrm{d}(\vartheta^+_1\psi^
+_1)^+_2+...]\nonumber \\
&&\qquad\qquad\qquad\qquad\qquad\quad\ \    +[\omega^-_2+\omega^+_3][\stackrel{\circ}{\vartheta}+\vartheta^+_1+\vartheta^-_2][\stackrel{\circ}{\psi}+\psi^
+_1+\psi^-_2]\} \nonumber \\
&&\quad\qquad\approx{\not\!\! X}(\textrm{d}\vartheta^-_2+\textrm{d}\psi^-_2+\omega^+_3+...)
\end{eqnarray}
\hrulefill
\begin{eqnarray}\label{fot3}
&&\quad\qquad\textrm{D}(\overline{\psi}{\not\!\! X}\textrm{i}\gamma)\wedge(\vartheta\psi)\approx(\overline{\psi}\vartheta)\wedge \textrm{D}(\textrm{i}\gamma{\not\!\! X}\psi) \nonumber \\
&&\quad\qquad\approx[\stackrel{\circ}{\overline{\psi}}+\overline{\psi}^+_1+\overline{\psi}^-_2][\stackrel{\circ}{\vartheta}+\vartheta^
+_1+\vartheta^-_2]\{[\textrm{d}{\not\!\! X}(\stackrel{\circ}{\psi}+\psi^
+_1+\psi^-_2)+{\not\!\! X}(\textrm{d}\psi^
+_1+\textrm{d}\psi^-_2)]\nonumber \\
&&\qquad\qquad\qquad\qquad\qquad\qquad\qquad\qquad\ \>\> +[\omega^-_2+\omega^+_3]{\not\!\! X}[\stackrel{\circ}{\overline{\psi}}+\overline{\psi}^+_1+\overline{\psi}^-_2]\} \nonumber \\
&&\quad\qquad\approx\textrm{d}{\not\!\! X}(\stackrel{\circ}{\overline{\psi}}+\stackrel{\circ}{\vartheta}+\stackrel{\circ}{\psi}+\overline{\psi}^-_2+\vartheta^-_2+\psi^-_2+...)+{\not\!\! X}(\textrm{d}\psi^-_2+\omega^+_3+...)
\end{eqnarray}
\hrulefill
\begin{eqnarray}\label{fot4}
&&\quad\qquad\textrm{D}\overline{\psi}\wedge (\vartheta \textrm{i}\gamma {\not\!\! X}\psi)\approx(\overline{\psi}{\not\!\! X}\textrm{i}\gamma\vartheta)\wedge \textrm{D}\psi\nonumber \\
&&\quad\qquad \approx[\stackrel{\circ}{\overline{\psi}}+\overline{\psi}^+_1+\overline{\psi}^-_2]{\not\!\! X}[\stackrel{\circ}{\vartheta}+\vartheta^
+_1+\vartheta^-_2]\{[\textrm{d}\psi^+_1+\textrm{d}\psi^-_2] \nonumber \\
&&\qquad\qquad\qquad\qquad\qquad\qquad\qquad\quad\quad\quad\ \>\,+[\omega^-_2+\omega^+_3][\stackrel{\circ}{\psi}+\psi^
+_1+\psi^-_2]\}\nonumber \\
&&\quad\qquad\approx{\not\!\! X}(\textrm{d}\psi^-_2+\omega^+_3+...)
\end{eqnarray}
In the calculation of the asymptotic falloffs in (\ref{fot1}), (\ref{fot2}), (\ref{fot3}) and (\ref{fot4}), we only illustrate those terms which will make contributions when integrated asymptotically in the final step and use ellipses to represent those terms which won't contribute, namely those terms having asymptotic falloffs faster than $\rm{O}(\frac{1}{r^2})$ and those having even parities. We see in the four calculations, all of these terms contribute a finite but nonvanishing value when integrated over the boundary bordering an asymptotically flat spatial 3-region. That is good, as we expect.

In the following let us inspect the form to which the Hamiltonian boundary integrand will approach asymptotically. We can choose any among the infinite number of expressions to inspect. Let us just choose the expression arrived at by adding
\begin{eqnarray}
&&\qquad -\textrm{d}[\overline{\psi}(\vartheta \textrm{i}\gamma{\not\!\! X}-{\not\!\! X}\textrm{i}\gamma\vartheta)\psi] \nonumber \\
&&\qquad=\frac{1}{2}\{[-\textrm{D}(\overline{\psi}\vartheta) (\textrm{i}\gamma {\not\!\! X}\psi)+(\overline{\psi}\vartheta)\wedge\textrm{D}(\textrm{i}\gamma{\not\!\! X}\psi)+\textrm{D}(\overline{\psi}{\not\!\! X}\textrm{i}\gamma)\wedge(\vartheta\psi)+(\overline{\psi}{\not\!\! X}\textrm{i}\gamma) \textrm{D}(\vartheta\psi)]
\nonumber \\
&&\qquad\qquad\ \> +[-\textrm{D}\overline{\psi}\wedge(\vartheta \textrm{i}\gamma {\not\!\! X}\psi)-\overline{\psi}\textrm{D}(\vartheta \textrm{i}\gamma{\not\!\! X}\psi)+\textrm{D}(\overline{\psi}{\not\!\! X}\textrm{i}\gamma\vartheta)\psi-(\overline{\psi}{\not\!\! X}\textrm{i}\gamma\vartheta)\wedge \textrm{D}\psi]\}
\end{eqnarray}
to the boundary expression in (\ref{HA4}).
\begin{eqnarray}\label{4-termboundary}
&&\qquad -\overline{\psi}\textrm{\textrm{D}}(\vartheta \textrm{i}\gamma{\not\!\! X}\psi)+\textrm{\textrm{D}}(\overline{\psi}{\not\!\! X}\textrm{i}\gamma\vartheta)\psi  +(\overline{\psi}\vartheta)\wedge \textrm{\textrm{D}}(\textrm{i}\gamma{\not\!\! X}\psi)+\textrm{\textrm{D}}(\overline{\psi}{\not\!\! X}\textrm{i}\gamma)\wedge(\vartheta\psi) \nonumber \\
&&\qquad=-\overline{\psi}\stackrel{\circ}{\textrm{D}}(\vartheta \textrm{i}\gamma{\not\!\! X}\psi)+\stackrel{\circ}{\textrm{D}}(\overline{\psi}{\not\!\! X}\textrm{i}\gamma\vartheta)\psi  +(\overline{\psi}\vartheta)\wedge \stackrel{\circ}{\textrm{D}}(\textrm{i}\gamma{\not\!\! X}\psi)+\stackrel{\circ}{\textrm{D}}(\overline{\psi}{\not\!\! X}\textrm{i}\gamma)\wedge(\vartheta\psi) \nonumber \\
&&\qquad\quad-\overline{\psi}\Delta\omega\wedge(\vartheta \textrm{i}\gamma{\not\!\! X}\psi)+(\overline{\psi}{\not\!\! X}\textrm{i}\gamma\vartheta)\wedge\Delta\omega\psi  \nonumber \\
&&\qquad\quad+(\overline{\psi}\vartheta)\wedge \Delta\omega( \textrm{i}\gamma{\not\!\! X}\psi)-(\overline{\psi}{\not\!\! X}\textrm{i}\gamma)\Delta\omega\wedge(\vartheta\psi) \nonumber \\
&&\qquad=\quad\ \textrm{d}[-\stackrel{\circ}{\overline{\psi}}(\vartheta \textrm{i}\gamma{\not\!\! X}\psi)+(\overline{\psi}{\not\!\! X}\textrm{i}\gamma\vartheta)\stackrel{\circ}{\psi}  -\stackrel{\circ}{(\overline{\psi}\vartheta)}(\textrm{i}\gamma{\not\!\! X}\psi)+(\overline{\psi}{\not\!\! X}\textrm{i}\gamma)\stackrel{\circ}{(\vartheta\psi)}] \nonumber \\
&&\qquad\quad\ -\Delta\overline{\psi}\stackrel{\circ}{\textrm{D}}(\vartheta \textrm{i}\gamma{\not\!\! X}\psi)+\stackrel{\circ}{\textrm{D}}(\overline{\psi}{\not\!\! X}\textrm{i}\gamma\vartheta)\Delta\psi  \nonumber \\
&&\qquad\quad\ +\Delta(\overline{\psi}\vartheta)\wedge \stackrel{\circ}{\textrm{D}}(\textrm{i}\gamma{\not\!\! X}\psi) +\stackrel{\circ}{\textrm{D}}(\overline{\psi}{\not\!\! X}\textrm{i}\gamma)\wedge\Delta(\vartheta\psi)\nonumber \\
&&\qquad\quad\ -\frac{\textrm{1}}{4}\Delta\omega^{\alpha\beta}\wedge\vartheta^\lambda\overline{\psi}[(\gamma_{\alpha\beta}\gamma_\lambda+\gamma_\lambda\gamma_{\alpha\beta})\textrm{i}\gamma {\not\!\! X}+{\not\!\! X}\textrm{i}\gamma(\gamma_\lambda\gamma_{\alpha\beta}+\gamma_{\alpha\beta}\gamma_\lambda)]\psi
\nonumber \\
&&\qquad=\quad\ \textrm{d}[-\stackrel{\circ}{\overline{\psi}}(\vartheta \textrm{i}\gamma{\not\!\! X}\psi)+(\overline{\psi}{\not\!\! X}\textrm{i}\gamma\vartheta)\stackrel{\circ}{\psi}  -\stackrel{\circ}{(\overline{\psi}\vartheta)} (\textrm{i}\gamma{\not\!\! X}\psi)+(\overline{\psi}{\not\!\! X}\textrm{i}\gamma)\stackrel{\circ}{(\vartheta\psi)}] \nonumber \\
&&\qquad\quad\ -\Delta\overline{\psi}\stackrel{\circ}{\textrm{D}}(\vartheta \textrm{i}\gamma{\not\!\! X}\psi)+\stackrel{\circ}{\textrm{D}}(\overline{\psi}{\not\!\! X}\textrm{i}\gamma\vartheta)\Delta\psi  \nonumber \\
&&\qquad\quad\ +\Delta(\overline{\psi}\vartheta)\wedge \stackrel{\circ}{\textrm{D}}(\textrm{i}\gamma{\not\!\! X}\psi)+\stackrel{\circ}{\textrm{D}}(\overline{\psi}{\not\!\! X}\textrm{i}\gamma)\wedge\Delta(\vartheta\psi) \nonumber \\
&&\qquad\quad\ -\Delta\omega^{\alpha\beta}\wedge\eta_{\alpha\beta\kappa}\frac{\textrm{i}}{2}\overline{\psi}[\gamma^\kappa{\not\!\! X}-
{\not\!\! X}\gamma^\kappa]\psi
\end{eqnarray}
In the above, we see the fourth line of the second step contains the element in the left-hand side of (\ref{repetitivepattern}), so we just invoke the result gained there.
In the first line of the second step, we pull out $\stackrel{\circ}{\textrm{D}}$ to become $\textrm{d}$ for $\stackrel{\circ}{\textrm{D}}\stackrel{\circ}{\overline{\psi}}=\stackrel{\circ}{\textrm{D}}\stackrel{\circ}{\psi}
=\stackrel{\circ}{\textrm{D}}\stackrel{\circ}{(\overline{\psi}\vartheta)}=\stackrel{\circ}{\textrm{D}}\stackrel{\circ}{(\vartheta\psi)}=0$ and when this line is inserted back into the Hamiltonian 3-form, we have $\textrm{dd}(\overline{\psi}\textsf{A}\wedge \textsf{B}\psi)=0$, some terms being therefore canceled. If we choose the other boundary expressions, this step doesn't hold because $\stackrel{\circ}{\textrm{D}}\stackrel{\circ}{(\overline{\psi}\textsf{A})}$ or $\stackrel{\circ}{\textrm{D}}\stackrel{\circ}{(\textsf{B}\psi)}$ doesn't vanish when $\textsf{A}$ or $\textsf{B}$ contains ${\not\!\! X}$.
On the other hand, the final line is the asymptotical form which we hope our Hamiltonian boundary integrand approaches. So we hope the other terms in the final step will vanish when integrated asymptotically. Let us first check if this desired form gives a finite value when integrated asymptotically with our choice of the asymptotic field falloffs with parities. Because we choose the background geometry to be Minkowski space, the reference value of the connection $\stackrel{\circ}{\omega}$ can be chosen to vanish everywhere \footnote{Choosing $\stackrel{\circ}{\omega}=\textrm{cost}$ would make both the Hamiltonian boundary term and the Hamiltonian variation boundary term give infinite asymptotic integrals.} and thus we have $\omega\approx\Delta\omega$.
\begin{eqnarray}\label{foADMB}
&&\qquad\Delta\omega^{\alpha\beta}\wedge\eta_{\alpha\beta\kappa}\frac{\textrm{i}}{2}\overline{\psi}[\gamma^\kappa{\not\!\! X}-
{\not\!\! X}\gamma^\kappa]\psi \nonumber \\
&&\qquad\approx [{\Delta\omega^{\alpha\beta}}^-_2+{\Delta\omega^{\alpha\beta}}^+_3] [\stackrel{\circ}{(\eta_{\alpha\beta\kappa})}+\eta_{\alpha\beta\kappa}{}^+_1+\eta_{\alpha\beta\kappa}{}^-_2][\stackrel{\circ}{\overline{\psi}}+\overline{\psi}^+_1+\overline{\psi}^-_2]{\not\!\! X}[\stackrel{\circ}{\psi}+\psi^+_1+\psi^-_2] \nonumber\\
&&\qquad\approx({\Delta\omega^{\alpha\beta}}^+_3{\not\!\! X})^-_2+...
\end{eqnarray}
We see that it adapts to our requirement, giving a nonvanishing finite integral value asymptotically. Then let us see if the other terms of the Hamiltonian boundary 2-form each give a vanishing asymptotic integral. From the process of (\ref{fot1}) and (\ref{fot3}), we find
\begin{eqnarray}\label{HB3}
\Delta\overline{\psi}\stackrel{\circ}{\textrm{D}}(\vartheta \textrm{i}\gamma{\not\!\! X}\psi)\approx\stackrel{\circ}{\textrm{D}}(\overline{\psi}{\not\!\! X}\textrm{i}\gamma\vartheta)\Delta\psi\approx \textrm{d}{\not\!\! X}\Delta\psi^-_2+...
\end{eqnarray}
for the second line and
\begin{eqnarray}\label{HB4}
\stackrel{\circ}{\textrm{D}}(\overline{\psi}{\not\!\! X}\textrm{i}\gamma)\wedge\Delta(\vartheta\psi)\approx\Delta(\overline{\psi}\vartheta )\wedge \stackrel{\circ}{\textrm{D}}(\textrm{i}\gamma{\not\!\! X}\psi)
\approx(\Delta\overline{\psi}^-_2+\Delta\vartheta^-_2+...)\textrm{d}{\not\!\! X}
\end{eqnarray}
for the third line of the final step. It's bad that we don't get what we desire; the second and third lines of the final step contribute a nonvanishing asymptotic integral. It can be shown that if the other boundary expressions were chosen, the terms contributing to the asymptotic integral would be the same, even though pulling out $\stackrel{\circ}{\textrm{D}}$ to become $\textrm{d}$ in the second step of (\ref{4-termboundary}) doesn't hold for them. Thus when integrated asymptotically, the Hamiltonian boundary integral has the contribution from the redundant terms in addition to the desired ADM term.

In order to get the dynamical evolution equations, we need to vary the Hamiltonian. Let us apply the recipe
\begin{eqnarray}\label{variationrule}
&&\delta[\textrm{D}(\overline{\psi}\textsf{A})\wedge\textrm{D}(\textsf{B}\psi)] \nonumber \\
&&=\textrm{D}\delta(\overline{\psi}\textsf{A})\wedge\textrm{D}(\textsf{B}\psi)
+\textrm{D}(\overline{\psi}\textsf{A})\wedge\textrm{D}\delta(\textsf{B}\psi) \nonumber \\
&&\quad-(-1)^a(\overline{\psi}\textsf{A})\wedge\delta\omega\wedge\textrm{D}(\textsf{B}\psi)+\textrm{D}(\overline{\psi}\textsf{A})\wedge\delta\omega\wedge(\textsf{B}\psi) \nonumber \\
&&=\textrm{d}[\delta(\overline{\psi}\textsf{A})\wedge\textrm{D}(\textsf{B}\psi)-(-1)^a\textrm{D}(\overline{\psi}\textsf{A})\wedge\delta(\textsf{B}\psi)] \nonumber \\
&&\quad-(-1)^a\delta(\overline{\psi}\textsf{A})\wedge\textrm{D}^2(\textsf{B}\psi)+(-1)^a\textrm{D}^2(\overline{\psi}\textsf{A})\wedge\delta(\textsf{B}\psi) \nonumber \\
&&\quad+\frac{1}{4}\delta\omega^{\alpha\beta}\wedge[-(\overline{\psi}\textsf{A})\gamma_{\alpha\beta}\wedge\textrm{D}(\textsf{B}\psi)
-(-1)^a\textrm{D}(\overline{\psi}\textsf{A})\gamma_{\alpha\beta}\wedge(\textsf{B}\psi)] \nonumber \\
&&=\textrm{d}[\delta(\overline{\psi}\textsf{A})\wedge\stackrel{\circ}{\textrm{D}}(\textsf{B}\psi)-(-1)^a\stackrel{\circ}{\textrm{D}}(\overline{\psi}\textsf{A})\wedge\delta(\textsf{B}\psi) \nonumber \\
&&\quad\quad+\delta(\overline{\psi}\textsf{A})\wedge\Delta\omega\wedge(\textsf{B}\psi)+(\overline{\psi}\textsf{A})\wedge\Delta\omega\wedge\delta(\textsf{B}\psi)] \nonumber \\
&&\quad-(-1)^a\delta(\overline{\psi}\textsf{A})\wedge \textsf{R}\wedge(\textsf{B}\psi)-(-1)^a(\overline{\psi}\textsf{A})\wedge \textsf{R}\wedge\delta(\textsf{B}\psi) \nonumber \\
&&\quad+\frac{1}{4}\delta\omega^{\alpha\beta}\wedge[-(\overline{\psi}\textsf{A})\gamma_{\alpha\beta}\wedge\textrm{D}(\textsf{B}\psi)
-(-1)^a\textrm{D}(\overline{\psi}\textsf{A})\gamma_{\alpha\beta}\wedge(\textsf{B}\psi)] \nonumber \\
&&=\quad\textrm{d}[\delta(\overline{\psi}\textsf{A})\wedge\stackrel{\circ}{\textrm{D}}(\textsf{B}\psi)-(-1)^a\stackrel{\circ}{\textrm{D}}(\overline{\psi}\textsf{A})\wedge\delta(\textsf{B}\psi)] \nonumber \\
&&\quad\,\,+\textrm{d}\{\frac{1}{4}\Delta\omega^{\alpha\beta}\wedge\delta[(-1)^a(\overline{\psi}\textsf{A})\gamma_{\alpha\beta}\wedge(\textsf{B}\psi)]\} \nonumber \\
&&\quad\,\,+\frac{1}{4}R^{\alpha\beta}\wedge\delta[-(-1)^a(\overline{\psi}\textsf{A})\gamma_{\alpha\beta}\wedge(\textsf{B}\psi)] \nonumber \\
&&\quad\,\,+\frac{1}{4}\delta\omega^{\alpha\beta}\wedge\textrm{D}[-(-1)^a(\overline{\psi}\textsf{A})\gamma_{\alpha\beta}\wedge(\textsf{B}\psi)]
\end{eqnarray}
to the variation of each of the four terms in the first and second lines of (\ref{HA4}).
For what inside the square brackets of the first line in the final step of (\ref{variationrule}), we have
\begin{eqnarray}
&&\quad-\delta\overline{\psi} \stackrel{\circ}{\textrm{D}}(\vartheta \textrm{i}\gamma {\not\!\! X}\psi)+\delta(\overline{\psi}{\not\!\! X} \textrm{i}\gamma\vartheta)\wedge \stackrel{\circ}{\textrm{D}}\psi-\delta(\overline{\psi}{\not\!\! X}\textrm{i}\gamma) \stackrel{\circ}{\textrm{D}}(\vartheta\psi)+\delta(\overline{\psi}\vartheta)\wedge \stackrel{\circ}{\textrm{D}}(\textrm{i}\gamma{\not\!\! X}\psi) \nonumber \\
&&\quad+\stackrel{\circ}{\textrm{D}}\overline{\psi}\wedge \delta(\vartheta \textrm{i}\gamma {\not\!\! X}\psi)+\stackrel{\circ}{\textrm{D}}(\overline{\psi}{\not\!\! X} \textrm{i}\gamma\vartheta) \delta\psi+\stackrel{\circ}{\textrm{D}}(\overline{\psi}{\not\!\! X}\textrm{i}\gamma)\wedge \delta(\vartheta\psi)+\stackrel{\circ}{\textrm{D}}(\overline{\psi}\vartheta) \delta(\textrm{i}\gamma{\not\!\! X}\psi) \nonumber \\
\end{eqnarray}
for the sum of the four terms.
And we see the second, third and fourth lines contain the common element $(-1)^a(\overline{\psi}\textsf{A})\gamma_{\alpha\beta}\wedge(\textsf{B}\psi)$, the sum of which over the four terms is
\begin{eqnarray}
&&-\overline{\psi}\gamma_{\alpha\beta}(\vartheta \textrm{i}\gamma {\not\!\! X}\psi)-(\overline{\psi}{\not\!\! X} \textrm{i}\gamma\vartheta) \gamma_{\alpha\beta}\psi-(\overline{\psi}{\not\!\! X}\textrm{i}\gamma) \gamma_{\alpha\beta}(\vartheta\psi)-(\overline{\psi}\vartheta) \gamma_{\alpha\beta}(\textrm{i}\gamma{\not\!\! X}\psi)  \nonumber \\
&&=-\overline{\psi}[(\gamma_{\alpha\beta}\gamma_\lambda+\gamma_\lambda\gamma_{\alpha\beta})\textrm{i}\gamma{\not\!\! X}+{\not\!\! X} \textrm{i}\gamma(\gamma_\lambda\gamma_{\alpha\beta}+\gamma_{\alpha\beta}\gamma_\lambda)]\psi\vartheta^\lambda  \nonumber \\ &&=-2\textrm{i}\overline{\psi}(\gamma^\kappa{\not\!\! X}-{\not\!\! X}\gamma^\kappa)\psi\eta_{\alpha\beta\kappa}.
\end{eqnarray}
In the above, we invoke (\ref{repetitivepattern}) again.
Altogether we arrive at
\begin{eqnarray}\label{VSHAM}
&&\quad\delta\mathcal{H}(\psi, {\not\!\! X}) \nonumber \\
&&\quad=\delta[-\textrm{D}\overline{\psi}\wedge \textrm{D}(\vartheta \textrm{i}\gamma {\not\!\! X}\psi)+\textrm{D}(\overline{\psi}{\not\!\! X} \textrm{i}\gamma\vartheta)\wedge \textrm{D}\psi   \nonumber \\
&&\qquad\ \ \>-\textrm{D}(\overline{\psi}{\not\!\! X}\textrm{i}\gamma)\wedge \textrm{D}(\vartheta\psi)+\textrm{D}(\overline{\psi}\vartheta)\wedge \textrm{D}(\textrm{i}\gamma{\not\!\! X}\psi)] \nonumber \\
&&\quad=\ \ \textrm{d}[-\delta\overline{\psi} \stackrel{\circ}{\textrm{D}}(\vartheta \textrm{i}\gamma {\not\!\! X}\psi)+\delta(\overline{\psi}{\not\!\! X} \textrm{i}\gamma\vartheta)\wedge \stackrel{\circ}{\textrm{D}}\psi \nonumber \\
&&\qquad\quad\ \> -\delta(\overline{\psi}{\not\!\! X}\textrm{i}\gamma) \stackrel{\circ}{\textrm{D}}(\vartheta\psi)+\delta(\overline{\psi}\vartheta)\wedge \stackrel{\circ}{\textrm{D}}(\textrm{i}\gamma{\not\!\! X}\psi) \nonumber \\
&&\qquad\quad\ \>+\stackrel{\circ}{\textrm{D}}\overline{\psi}\wedge \delta(\vartheta \textrm{i}\gamma {\not\!\! X}\psi)+\stackrel{\circ}{\textrm{D}}(\overline{\psi}{\not\!\! X} \textrm{i}\gamma\vartheta) \delta\psi \nonumber \\
&&\qquad\quad\ \>+\stackrel{\circ}{\textrm{D}}(\overline{\psi}{\not\!\! X}\textrm{i}\gamma)\wedge \delta(\vartheta\psi)+\stackrel{\circ}{\textrm{D}}(\overline{\psi}\vartheta) \delta(\textrm{i}\gamma{\not\!\! X}\psi)]
\nonumber \\
&&\qquad-\textrm{d}\{\Delta\omega^{\alpha\beta}\wedge\delta
[\frac{\textrm{i}}{2}\overline{\psi}(\gamma^\kappa{\not\!\! X}-{\not\!\! X}\gamma^\kappa)\psi\eta_{\alpha\beta\kappa}]\}
\nonumber \\
&&\qquad+R^{\alpha\beta}\wedge\delta[\frac{\textrm{i}}{2}\overline{\psi}(\gamma^\kappa{\not\!\! X}-{\not\!\! X}\gamma^\kappa)\psi\eta_{\alpha\beta\kappa}]
\nonumber \\
&&\qquad+\delta\omega^{\alpha\beta}\wedge \textrm{D}[\frac{\textrm{i}}{2}\overline{\psi}(\gamma^\kappa{\not\!\! X}-{\not\!\! X}\gamma^\kappa)\psi\eta_{\alpha\beta\kappa}]
\end{eqnarray}

For the field equation, determined by the final two lines of (\ref{VSHAM}), to be well defined, the part of the boundary integral of the Hamiltonian variation associated with the variation of the dynamical variables needs to vanish.
We see that, in the Hamiltonian $\mathcal{H}(\psi, {\not\!\! X})$, $\eta_{\alpha\beta\kappa}$ (or equivalently $\vartheta^\lambda$) and $\omega^{\alpha\beta}$ are certainly the dynamical fields of a gravitational system. Is the spinor a dynamical field? In other words, does the spinor field contain any physical information of a gravitational system? At first glance, we see the spinor is introduced to parameterize the rotation displacement, which is an external choice, being a non-physical and non-dynamical variable. Thus the spinor should be a gauge field. Is it really the case? Let us scrutinize it in more detail.  We see a subtle point emerges:
\begin{eqnarray}
\textrm{i}_{N}\vartheta^\kappa=N^\kappa=\overline{\psi}\textrm{i}\gamma^\kappa{}_\nu\psi x^\nu.
\end{eqnarray}
The components of the displacement vector field appear from the interior product of the coframe field and the components are parameterized by the spinor field. Thus the spinor contains the physical information involved in $\vartheta^\kappa$. Then we need to require the part of the  asymptotic boundary integral of the Hamiltonian variation involved with the spinor field, along with $\vartheta^\lambda$, to vanish.
Let us see if it vanishes.
Because the falloffs with parities of $\delta(\textsf{B}\psi)$ is the same as $\Delta(\textsf{B}\psi)$, we find
\begin{equation}\label{fpbhv1}
\delta\overline{\psi} \stackrel{\circ}{\textrm{D}}(\vartheta \textrm{i}\gamma {\not\!\! X}\psi)\approx\stackrel{\circ}{\textrm{D}}(\overline{\psi}{\not\!\! X} \textrm{i}\gamma\vartheta) \delta\psi\approx \textrm{d}{\not\!\! X}\delta\psi^-_2+...
\end{equation}
from (\ref{HB3}),
\begin{eqnarray}\label{fpbhv2}
&&\delta(\overline{\psi}{\not\!\! X} \textrm{i}\gamma\vartheta)\wedge \stackrel{\circ}{\textrm{D}}\psi\approx\stackrel{\circ}{\textrm{D}}\overline{\psi}\wedge \delta(\vartheta \textrm{i}\gamma {\not\!\! X}\psi) \approx \textrm{vanishing integral asymptotically} \nonumber \\
\end{eqnarray}
from the process of (\ref{fot4}) and that the fact that we don't vary $x$ because $x$ is not a dynamical field,
\begin{eqnarray}\label{fpbhv3}
&&\quad x^\nu\delta(\overline{\psi}\gamma_\nu \textrm{i}\gamma) \stackrel{\circ}{\textrm{D}}(\vartheta\psi)\approx\stackrel{\circ}{\textrm{D}}(\overline{\psi}\vartheta) x^\nu\delta(\textrm{i}\gamma\gamma_\nu\psi)
\approx \textrm{vanishing integral asymptotically}  \nonumber \\
\end{eqnarray}
from the process of (\ref{fot2}),
\begin{eqnarray}\label{fpbhv4}
&&\delta(\overline{\psi}\vartheta)\wedge \stackrel{\circ}{\textrm{D}}(\textrm{i}\gamma{\not\!\! X}\psi)\approx\stackrel{\circ}{\textrm{D}}(\overline{\psi}{\not\!\! X}\textrm{i}\gamma)\wedge \delta(\vartheta\psi)\approx\textrm{d}{\not\!\! X}(\delta\vartheta^-_2+\delta\psi^-_2+...)
\end{eqnarray}
from (\ref{HB4}), and
\begin{eqnarray}\label{fpbhv5}
&&\Delta\omega^{\alpha\beta}x^\nu\wedge\delta
[\frac{\textrm{i}}{2}\overline{\psi}(\gamma^\kappa\gamma_\nu-\gamma_\nu\gamma^\kappa)\psi\eta_{\alpha\beta\kappa}]
\approx\textrm{vanishing integral asymptotically} \nonumber \\
\end{eqnarray}
from the process of (\ref{foADMB}). Because the part of the asymptotic boundary integral of the Hamiltonian variation associated with the dynamical variables doesn't vanish, the field equations are not well defined.

In conclusion, the Hamiltonian 3-form $\mathcal{H}(\psi, {\not\!\! X})$ fails to be an expression for the angular momentum because its boundary integral does not approach the integral of the ADM term asymptotically and the part of the asymptotical boundary integral of its variation involved with the dynamical variable variation does not vanish with our choice of the asymptotic field falloffs with parities.

\section{Witten spinor Hamiltonian adapted for angular momentum with the bivectorial parameterization, the 2-term among the 4-term}\label{2-term}

From the process of the trial in the previous section, we found there is impossibility of a choice of the asymptotic field falloffs with parities which makes the part of the asymptotical boundary integral of the Hamiltonian variation associated with the variation of the dynamical variables vanish as well as making the Hamiltonian boundary 2-form have a finite but nonvanishing asymptotic integral value approaching the asymptotic integral of the ADM term. But we guess it might be successful if we choose different asymptotic falloffs with parities for the spinor field and hold only the promising 2 terms among the full 4-term quadratic spinor Hamiltonian (\ref{HA4}).

We consider the spinor-curvature identity
\begin{eqnarray}
&&\qquad -\textrm{D}\overline{\psi}\wedge \textrm{D}(\vartheta \textrm{i}\gamma {\not\!\! X}\psi)+\textrm{D}(\overline{\psi}{\not\!\! X} \textrm{i}\gamma\vartheta)\wedge \textrm{D}\psi \nonumber \\
&&\qquad\equiv\overline{\psi}\textsf{R}\wedge (\vartheta \textrm{i}\gamma {\not\!\! X}\psi)+(\overline{\psi}{\not\!\! X} \textrm{i}\gamma\vartheta)\wedge \textsf{R}\psi, \nonumber \\
&&\qquad\quad+\frac{1}{2}\textrm{d}[-\overline{\psi}\textrm{D}(\vartheta \textrm{i}\gamma{\not\!\! X}\psi)+\textrm{D}(\overline{\psi}{\not\!\! X}\textrm{i}\gamma\vartheta)\psi+(\overline{\psi}{\not\!\! X}\textrm{i}\gamma\vartheta)\wedge \textrm{D}\psi+\textrm{D}\overline{\psi}\wedge (\vartheta \textrm{i}\gamma {\not\!\! X}\psi)],  \nonumber \\
\end{eqnarray}
among which
\begin{eqnarray}\label{SCISH1}
&&\qquad\overline{\psi}\textsf{R}\wedge (\vartheta \textrm{i}\gamma {\not\!\! X}\psi)+(\overline{\psi}{\not\!\! X} \textrm{i}\gamma\vartheta)\wedge \textsf{R}\psi, \nonumber \\
&&\qquad=\frac{1}{4}R^{\alpha\beta}\wedge\vartheta^\lambda\overline{\psi}(\gamma_{\alpha\beta}\gamma_\lambda\textrm{i}\gamma{\not\!\! X}+{\not\!\! X}\textrm{i}\gamma\gamma_\lambda
\gamma_{\alpha\beta})\psi \nonumber \\
&&\qquad=\frac{1}{4}R^{\alpha\beta}\wedge\vartheta^\lambda\overline{\psi}[(\gamma_{\alpha\beta\lambda}-g_{\alpha\lambda}\gamma_\beta+
g_{\beta\lambda}\gamma_\alpha)\textrm{i}\gamma{\not\!\! X}+{\not\!\! X}\textrm{i}\gamma(\gamma_{\lambda\alpha\beta}+g_{\lambda\alpha}\gamma_\beta
-g_{\lambda\beta}\gamma_\alpha)]\psi \nonumber\\
&&\qquad=-\frac{1}{4}R^{\alpha\beta}\wedge\vartheta^\lambda\overline{\psi}[\eta_{\alpha\beta\lambda\kappa}\gamma\gamma^\kappa\textrm{i}\gamma{\not\!\! X}+{\not\!\! X}
\textrm{i}\gamma\eta_{\lambda\alpha\beta\kappa}\gamma\gamma^\kappa]\psi \nonumber \\
&&\qquad\quad +\frac{\textrm{i}}{4}R^{\beta\alpha}\wedge\vartheta_\alpha\overline{\psi}[\gamma_\beta\gamma{\not\!\! X}-{\not\!\! X}\gamma\gamma_\beta]\psi +\frac{\textrm{i}}{4}R^{\alpha\beta}\wedge\vartheta_\beta\overline{\psi}[\gamma_\alpha\gamma{\not\!\! X}-{\not\!\! X}\gamma\gamma_\alpha]\psi \nonumber \\
&&\qquad=R^{\alpha\beta}\wedge\eta_{\alpha\beta\kappa}\frac{\textrm{i}}{4}\overline{\psi}[\gamma^\kappa{\not\!\! X}-{\not\!\! X}\gamma^\kappa]\psi -R^\alpha{}_\beta\wedge\vartheta^\beta\frac{\textrm{i}}{2}\overline{\psi}[\gamma_\alpha{\not\!\! X}-{\not\!\! X}\gamma_\alpha]\gamma\psi.
\end{eqnarray}
Here we attain the result,
\begin{eqnarray}\label{repetitivepattern1}
&&\vartheta^\lambda\overline{\psi}(\gamma_{\alpha\beta}\gamma_\lambda\textrm{i}\gamma{\not\!\! X}+{\not\!\! X}\textrm{i}\gamma\gamma_\lambda
\gamma_{\alpha\beta})\psi \nonumber \\
&&=\eta_{\alpha\beta\kappa}\textrm{i}\overline{\psi}(\gamma^\kappa{\not\!\! X}-{\not\!\! X}\gamma^\kappa)\psi -2\vartheta_\beta\textrm{i}\overline{\psi}(\gamma_\alpha{\not\!\! X}-{\not\!\! X}\gamma_\alpha)\gamma\psi.
\end{eqnarray}

If we presume the vanishing torsion constraint, like the 4-term Hamiltonian, the second term of the final line of (\ref{SCISH1}) vanishes by the first Bianchi identity, $\textrm{d}\Theta^\alpha+\omega^\alpha{}_\beta\wedge\Theta^\beta=R^\alpha{}_\beta\wedge\vartheta^\beta$, and we are led to the Hamiltonian 3-form,
\begin{eqnarray}\label{HAf2}
&&\mathcal{H}_1(\psi,{\not\!\! X}):=2[-\textrm{D}\overline{\psi}\wedge \textrm{D}(\vartheta \textrm{i}\gamma {\not\!\! X}\psi)+\textrm{D}(\overline{\psi}{\not\!\! X} \textrm{i}\gamma\vartheta)\wedge \textrm{D}\psi] \nonumber \\
&&\qquad\qquad\>\;\equiv R^{\alpha\beta}\wedge\eta_{\alpha\beta\kappa}\frac{\textrm{i}}{2}\overline{\psi}[\gamma^\kappa{\not\!\! X}-{\not\!\! X}\gamma^\kappa]\psi \nonumber \\
&&\qquad\qquad\quad\>\;+\textrm{d}[-\overline{\psi}\textrm{D}(\vartheta \textrm{i}\gamma{\not\!\! X}\psi)+\textrm{D}(\overline{\psi}{\not\!\! X}\textrm{i}\gamma\vartheta)\psi \nonumber \\
&&\qquad\qquad\qquad\quad +(\overline{\psi}{\not\!\! X}\textrm{i}\gamma\vartheta)\wedge \textrm{D}\psi+\textrm{D}\overline{\psi}\wedge (\vartheta \textrm{i}\gamma {\not\!\! X}\psi)].
\end{eqnarray}
The Hamiltonian 3-form has the expected volume term.

Let us examine its boundary 2-form. We try to choose the asymptotic falloffs with parities of the spinor field as
\begin{eqnarray}
\overline{\psi}\approx\psi\approx\stackrel{\circ}{\psi}+\psi^+_1+\psi^+_2,
\end{eqnarray}
and keep the asymptotic falloffs with parities of the other fields the same as those for $\mathcal{H}(\psi, {\not\!\! X})$. We see the asymptotic boundary integral is finite but nonvanishing due to the contribution of ${\not\!\! X}\textrm{d}\vartheta^-_2$, $\textrm{d}{\not\!\! X}(\stackrel{\circ}{\overline{\psi}}\stackrel{\circ}{\vartheta}\stackrel{\circ}{\psi}+\vartheta^-_2)$ and $(\omega^+_3{\not\!\! X})^-_2$ from (\ref{fot1}) and (\ref{fot4}); the spinor of asymptotic falloff $\rm{O}(\frac{1}{r^2})$ no longer makes a contribution because we have changed its parity from odd to even in this section.

Let us check the asymptotic integral which its boundary integral approaches.
\begin{eqnarray}\label{HAP2}
&&\qquad[-\overline{\psi}\textrm{D}(\vartheta \textrm{i}\gamma{\not\!\! X}\psi)+\textrm{D}(\overline{\psi}{\not\!\! X}\textrm{i}\gamma\vartheta)\psi+(\overline{\psi}{\not\!\! X}\textrm{i}\gamma\vartheta)\wedge \textrm{D}\psi+\textrm{D}\overline{\psi}\wedge (\vartheta \textrm{i}\gamma {\not\!\! X}\psi)] \nonumber \\
&&\qquad=2[\textrm{D}\overline{\psi}\wedge (\vartheta \textrm{i}\gamma {\not\!\! X}\psi)+(\overline{\psi}{\not\!\! X}\textrm{i}\gamma\vartheta)\wedge \textrm{D}\psi] \nonumber \\
&&\qquad=\quad2[\stackrel{\circ}{\textrm{D}}\overline{\psi}\wedge (\vartheta \textrm{i}\gamma {\not\!\! X}\psi)+(\overline{\psi}{\not\!\! X}\textrm{i}\gamma\vartheta)\wedge \stackrel{\circ}{\textrm{D}}\psi] \nonumber \\
&&\quad\quad\quad\>+2[-\overline{\psi}\Delta\omega\wedge(\vartheta \textrm{i}\gamma{\not\!\! X}\psi)+(\overline{\psi}{\not\!\! X}\textrm{i}\gamma\vartheta)\wedge\Delta\omega\psi] \nonumber \\
&&\qquad=\quad2[\stackrel{\circ}{\textrm{D}}\overline{\psi}\wedge (\vartheta \textrm{i}\gamma {\not\!\! X}\psi)+(\overline{\psi}{\not\!\! X}\textrm{i}\gamma\vartheta)\wedge \stackrel{\circ}{\textrm{D}}\psi] \nonumber \\ &&\quad\quad\quad\>-\frac{1}{2}\Delta\omega^{\alpha\beta}\wedge\vartheta^\lambda\overline{\psi}(\gamma_{\alpha\beta}\gamma_\lambda\textrm{i}\gamma{\not\!\! X}+{\not\!\! X}\textrm{i}\gamma
\gamma_\lambda\gamma_{\alpha\beta})\psi \nonumber \\
&&\qquad=\quad2[\stackrel{\circ}{\textrm{D}}\overline{\psi}\wedge (\vartheta \textrm{i}\gamma {\not\!\! X}\psi)+(\overline{\psi}{\not\!\! X}\textrm{i}\gamma\vartheta)\wedge \stackrel{\circ}{\textrm{D}}\psi] \nonumber \\
&&\quad\quad\quad\ -\Delta\omega^{\alpha\beta}\wedge\eta_{\alpha\beta\kappa}\frac{\textrm{i}}{2}\overline{\psi}(\gamma^\kappa{\not\!\! X}-{\not\!\! X}\gamma^\kappa)\psi +\Delta\omega^\alpha{}_\beta\wedge\vartheta^\beta\textrm{i}\overline{\psi}(\gamma_\alpha{\not\!\! X}
-{\not\!\! X}\gamma_\alpha)\gamma\psi. \nonumber \\
\end{eqnarray}
Here we deliberately choose those terms for which $\stackrel{\circ}{\textrm{D}}$ can't be pulled out to become $\textrm{d}$ to show choosing these terms doesn't change the result: the contributing elements, namely $\stackrel{\circ}{\overline{\psi}}\stackrel{\circ}{\vartheta}\textrm{d}{\not\!\! X}\stackrel{\circ}{\psi}$, $\textrm{d}\vartheta^-_2{\not\!\! X}$ and $\vartheta^-_2\textrm{d}{\not\!\! X}$ in $\overline{\psi}\stackrel{\circ}{\textrm{D}}(\vartheta \textrm{i}\gamma{\not\!\! X}\psi)$ as well as $\stackrel{\circ}{\overline{\psi}}\textrm{d}{\not\!\! X}\stackrel{\circ}{\vartheta}\stackrel{\circ}{\psi}$, $\textrm{d}{\not\!\! X}\vartheta^-_2$ and ${\not\!\! X}\textrm{d}\vartheta^-_2$ in $\stackrel{\circ}{\textrm{D}}(\overline{\psi}{\not\!\! X}\textrm{i}\gamma\vartheta)\psi$, of (\ref{4-termboundary}) were canceled by means of pulling out $\stackrel{\circ}{\textrm{D}}$ into $\textrm{d}$, but these elements originally don't make contributions in the current expression. This expression is reached by adding (\ref{adding1}) to the boundary expression of (\ref{HAf2}). We see the left-hand side of (\ref{repetitivepattern1}) appears in the second line of the third step, so we just invoke the result we gained there.
And the first term in the final line is the desired ADM term. We know it contributes a finite but nonvanishing integral asymptotically from (\ref{foADMB}).
Then let us see if the other terms give vanishing integrals asymptotically.
\begin{eqnarray}\label{fosh2F}
&&\qquad(\overline{\psi}{\not\!\! X}\textrm{i}\gamma\vartheta)\wedge \stackrel{\circ}{\textrm{D}}\psi\approx\stackrel{\circ}{\textrm{D}}\overline{\psi}\wedge (\vartheta \textrm{i}\gamma {\not\!\! X}\psi) \approx\textrm{vanishing integral asymptotically } \nonumber \\
\end{eqnarray}
from the process of (\ref{fot4}).
It is good that the first line of the final step gives a vanishing asymptotical integral. As for the second term of the final line, because $\Delta\omega^\alpha{}_\beta\wedge\vartheta^\beta \approx\Delta\omega^{\alpha\beta}\wedge\eta_{\alpha\beta\kappa}$, we see it also contributes a finite but nonvanishing asymptotical integral value. It's bad. We hope it could vanish. Let us think if we can figure out some way to make it vanish. It looks like we can choose the assumption
\begin{eqnarray}
\textrm{i}\overline{\psi}\gamma_{\beta\nu}\gamma\psi=0
\end{eqnarray}
to make this term vanish.
But after examining it in detail, one would find such an assumption is infeasible for $\textrm{i}\overline{\psi}\gamma_{jk}\gamma\psi=-\textrm{i}\overline{\psi}\epsilon^{jkl}\gamma_{0l}\psi$ and $\textrm{i}\overline{\psi}\gamma_{0l}\gamma\psi=\textrm{i}/2\overline{\psi}\epsilon^{jkl}\gamma_{jk}\psi$; thus this assumption would make the ADM term vanish likewise. On the other hand, this redundant term $\Delta\omega^\alpha{}_\beta\wedge\vartheta^\beta \textrm{i}\overline{\psi}(\gamma_\alpha{\not\!\! X}
-{\not\!\! X}\gamma_\alpha)\gamma\psi$ looks, in some way, like the redundant term $-2N^\nu\Delta\omega_{\nu\mu}\wedge\vartheta^\mu\overline{\psi}\gamma\psi$ in the QSL Hamiltonian boundary 2-form \cite{qslEM,QSLteleparallel}. That redundant term is eliminated by the assumption of $\overline{\psi}\gamma\psi=0$, where $\psi$ is a gauge field. Probably we can also introduce some gauge field to get rid of this undesired term here. Before a feasible way is figured out to rid of it, the Hamiltonian boundary integral does not approach the ADM integral asymptotically and thus can't give the angular momentum.

Let us vary $\mathcal{H}_1$. By applying (\ref{variationrule}) to vary each of the two terms of the first line in (\ref{HAf2}), we get
\begin{eqnarray}\label{varyH1}
&&\qquad\delta\mathcal{H}_1(\psi,{\not\!\! X}) \nonumber \\
&&\qquad=2\delta[-\textrm{D}\overline{\psi}\wedge \textrm{D}(\vartheta \textrm{i}\gamma {\not\!\! X}\psi)+\textrm{D}(\overline{\psi}{\not\!\! X} \textrm{i}\gamma\vartheta)\wedge \textrm{D}\psi] \nonumber \\
&&\qquad=\>\>2\textrm{d}[-\delta\overline{\psi} \stackrel{\circ}{\textrm{D}}(\vartheta \textrm{i}\gamma {\not\!\! X}\psi)+\delta(\overline{\psi}{\not\!\! X} \textrm{i}\gamma\vartheta)\wedge \stackrel{\circ}{\textrm{D}}\psi \nonumber \\
&&\qquad\qquad\>\>\ +\stackrel{\circ}{\textrm{D}}\overline{\psi}\wedge \delta(\vartheta \textrm{i}\gamma {\not\!\! X}\psi)+\stackrel{\circ}{\textrm{D}}(\overline{\psi}{\not\!\! X} \textrm{i}\gamma\vartheta) \delta\psi] \nonumber \\
&&\qquad\quad\ -\textrm{d}\{\frac{1}{2}\Delta\omega^{\alpha\beta}\wedge\delta[\overline{\psi}\gamma_{\alpha\beta}(\vartheta \textrm{i}\gamma{\not\!\! X}\psi)+(\overline{\psi}{\not\!\! X}\textrm{i}\gamma\vartheta)\gamma_{\alpha\beta}\psi]\} \nonumber \\
&&\qquad\quad\ +\frac{1}{2}R^{\alpha\beta}\wedge\delta[\overline{\psi}\gamma_{\alpha\beta}(\vartheta \textrm{i}\gamma{\not\!\! X}\psi)+(\overline{\psi}{\not\!\! X}\textrm{i}\gamma\vartheta)\gamma_{\alpha\beta}\psi] \nonumber \\
&&\qquad\quad\ +\frac{1}{2}\delta\omega^{\alpha\beta}\wedge \textrm{D}[\overline{\psi}\gamma_{\alpha\beta}(\vartheta \textrm{i}\gamma{\not\!\! X}\psi)+(\overline{\psi}{\not\!\! X}\textrm{i}\gamma\vartheta)\gamma_{\alpha\beta}\psi] \nonumber \\
&&\qquad=\>\>2\textrm{d}[-\delta\overline{\psi} \stackrel{\circ}{\textrm{D}}(\vartheta \textrm{i}\gamma {\not\!\! X}\psi)+\delta(\overline{\psi}{\not\!\! X} \textrm{i}\gamma\vartheta)\wedge \stackrel{\circ}{\textrm{D}}\psi \nonumber \\
&&\qquad\qquad\>\>\ +\stackrel{\circ}{\textrm{D}}\overline{\psi}\wedge \delta(\vartheta \textrm{i}\gamma {\not\!\! X}\psi)+\stackrel{\circ}{\textrm{D}}(\overline{\psi}{\not\!\! X} \textrm{i}\gamma\vartheta) \delta\psi] \nonumber \\
&&\qquad\quad\ -\textrm{d}\{\frac{1}{2}\Delta\omega^{\alpha\beta}\wedge\delta[\vartheta^\lambda\overline{\psi}(\gamma_{\alpha\beta}\gamma_\lambda \textrm{i}\gamma{\not\!\! X}+{\not\!\! X}\textrm{i}\gamma\gamma_\lambda\gamma_{\alpha\beta})\psi]\}\nonumber \\
&&\qquad\quad\ +\frac{1}{2}R^{\alpha\beta}\wedge\delta[\vartheta^\lambda\overline{\psi}(\gamma_{\alpha\beta}\gamma_\lambda \textrm{i}\gamma{\not\!\! X}+{\not\!\! X}\textrm{i}\gamma\gamma_\lambda\gamma_{\alpha\beta})\psi]\nonumber \\
&&\qquad\quad\ +\frac{1}{2}\delta\omega^{\alpha\beta}\wedge \textrm{D}[\vartheta^\lambda\overline{\psi}(\gamma_{\alpha\beta}\gamma_\lambda \textrm{i}\gamma{\not\!\! X}+{\not\!\! X}\textrm{i}\gamma\gamma_\lambda\gamma_{\alpha\beta})\psi]\nonumber \\
&&\qquad=\>\>2\textrm{d}[-\delta\overline{\psi} \stackrel{\circ}{\textrm{D}}(\vartheta \textrm{i}\gamma {\not\!\! X}\psi)+\delta(\overline{\psi}{\not\!\! X} \textrm{i}\gamma\vartheta)\wedge \stackrel{\circ}{\textrm{D}}\psi \nonumber \\
&&\qquad\qquad\>\>\ +\stackrel{\circ}{\textrm{D}}\overline{\psi}\wedge \delta(\vartheta \textrm{i}\gamma {\not\!\! X}\psi)+\stackrel{\circ}{\textrm{D}}(\overline{\psi}{\not\!\! X} \textrm{i}\gamma\vartheta) \delta\psi] \nonumber \\
&&\qquad\quad\ -\textrm{d}\{\Delta\omega^{\alpha\beta}\wedge\delta[\eta_{\alpha\beta\kappa}\frac{\textrm{i}}{2}\overline{\psi}(\gamma^\kappa{\not\!\! X}-{\not\!\! X}\gamma^\kappa)\psi] -\Delta\omega^\alpha{}_\beta\wedge\delta [\vartheta^\beta\textrm{i}\overline{\psi}(\gamma_\alpha{\not\!\! X}-{\not\!\! X}\gamma_\alpha)\gamma\psi]\}
\nonumber \\
&&\qquad\quad\ +R^{\alpha\beta}\wedge\delta[\eta_{\alpha\beta\kappa}\frac{\textrm{i}}{2}\overline{\psi}(\gamma^\kappa{\not\!\! X}-{\not\!\! X}\gamma^\kappa)\psi] -R^\alpha{}_\beta\wedge\delta [\vartheta^\beta \textrm{i}\overline{\psi}(\gamma_\alpha{\not\!\! X}-{\not\!\! X}\gamma_\alpha)\gamma\psi] \nonumber \\
&&\qquad\quad\ +\delta\omega^{\alpha\beta}\wedge \textrm{D}[\eta_{\alpha\beta\kappa}\frac{\textrm{i}}{2}\overline{\psi}(\gamma^\kappa{\not\!\! X}-{\not\!\! X}\gamma^\kappa)\psi]-\delta\omega^\alpha{}_\beta\wedge \textrm{D}[\vartheta^\beta \textrm{i}\overline{\psi}(\gamma_\alpha{\not\!\! X}-{\not\!\! X}\gamma_\alpha)\gamma\psi]. \nonumber \\
\end{eqnarray}
In the third step, we see the left-hand side of (\ref{repetitivepattern1}) reappears three times, so we just invoke the result we got there.
Let us see if the part of the Hamiltonian variation boundary term associated with the variation of $\vartheta^\lambda$, $\psi$ and $\overline{\psi}$ gives a vanishing asymptotical integral.
\begin{eqnarray}
&&\qquad\delta\overline{\psi}\wedge \stackrel{\circ}{\textrm{D}}(\vartheta \textrm{i}\gamma {\not\!\! X}\psi)\approx\stackrel{\circ}{\textrm{D}}(\overline{\psi}{\not\!\! X} \textrm{i}\gamma\vartheta)\wedge \delta\psi\approx \textrm{vanishing integral asymptotically} \nonumber \\
\end{eqnarray}
from (\ref{fpbhv1}).
\begin{eqnarray}
&&\qquad x^\nu\delta(\overline{\psi}\gamma_\nu \textrm{i}\gamma\vartheta)\wedge \stackrel{\circ}{\textrm{D}}\psi \approx \stackrel{\circ}{\textrm{D}}\overline{\psi}\wedge x^\nu\delta(\vartheta \textrm{i}\gamma \gamma_\nu\psi)\approx \textrm{vanishing integral asymptotically} \nonumber \\
\end{eqnarray}
from (\ref{fosh2F}).
\begin{eqnarray}\label{H2BC}
&&\qquad\Delta\omega^\alpha{}_\beta\wedge x^\nu\delta [\vartheta^\beta \textrm{i}\overline{\psi}(\gamma_\alpha\gamma_\nu-\gamma_\nu\gamma_\alpha)\gamma\psi] \approx\Delta\omega^{\alpha\beta}\wedge x^\nu\delta[\eta_{\alpha\beta\kappa}\frac{\textrm{i}}{2}\overline{\psi}(\gamma^\kappa\gamma_\nu-\gamma_\nu\gamma^\kappa)\psi] \nonumber \\
&&\qquad\approx \textrm{vanishing integral asymptotically}
\end{eqnarray}
from (\ref{fpbhv5}).
The asymptotic boundary integral of the Hamiltonian variation associated with the dynamical variable variation vanishes. Therefore the field equations are well defined. The dynamical variables are $\vartheta^\lambda$ (and part of $\overline{\psi}$ and $\psi$) and $\omega^{\alpha\beta}$. The boundary condition, manifested in the symplectic structure of the third line of the final step in (\ref{varyH1}), the first term of which is the expected boundary term of the Hamiltonian variation (\ref{varyHamiltonian}) when the torsion is assumed to vanish \textit{a priori}, is that we should control $\vartheta^\lambda$ (it will also in turn control $\overline{\psi}$ and $\psi$ partially) and leave $\omega^{\alpha\beta}$ as the response variable.

In conclusion, the only problem of $\mathcal{H}_1(\psi, {\not\!\! X})$ is that its asymptotical boundary 2-form contains a redundant term in addition to the ADM term. If we can find some way to remove it and yet preserve the desired terms, we can succeed in it.

\section{Witten spinor Hamiltonian adapted for angular momentum with the bivectorial parameterization, the other 2-term among the 4-term}\label{other2-term}

We may as well take a look at the spinor Hamiltonian composed of twice the other 2 terms among $\mathcal{H}(\psi, {\not\!\! X})$ which $\mathcal{H}_1(\psi, {\not\!\! X})$ doesn't contain with the asymptotic field falloffs with parities taken to be the same as those for $\mathcal{H}_1(\psi,{\not\!\! X})$.
\begin{eqnarray}\label{HAs2}
&&\mathcal{H}_2(\psi,{\not\!\! X}):=2[-\textrm{D}(\overline{\psi}{\not\!\! X}\textrm{i}\gamma)\wedge \textrm{D}(\vartheta\psi)+\textrm{D}(\overline{\psi}\vartheta)\wedge \textrm{D}(\textrm{i}\gamma {\not\!\! X}\psi)] \nonumber \\
&&\qquad\qquad\>\;\equiv 2[(\overline{\psi}{\not\!\! X}\textrm{i}\gamma)\textsf{R}\wedge (\vartheta\psi)+(\overline{\psi}\vartheta)\wedge \textsf{R} (\textrm{i}\gamma {\not\!\! X}\psi)] \nonumber \\
&&\qquad\qquad\quad\>\;+\textrm{d}[-(\overline{\psi}{\not\!\! X}\textrm{i}\gamma) \textrm{D}(\vartheta\psi)+\textrm{D}(\overline{\psi}\vartheta) (\textrm{i}\gamma {\not\!\! X}\psi) \nonumber \\
&&\qquad\qquad\qquad\quad+(\overline{\psi}\vartheta)\wedge\textrm{D}(\textrm{i}\gamma{\not\!\! X}\psi)+\textrm{D}(\overline{\psi}{\not\!\! X}\textrm{i}\gamma)\wedge(\vartheta\psi)] \nonumber \\
&&\qquad\qquad\quad=\frac{1}{2}R^{\alpha\beta}\wedge\vartheta^\lambda\overline{\psi}({\not\!\! X}\textrm{i}\gamma\gamma_{\alpha\beta}\gamma_\lambda+\gamma_\lambda
\gamma_{\alpha\beta}\textrm{i}\gamma{\not\!\! X})\psi \nonumber \\
&&\qquad\qquad\qquad+\textrm{d}[-(\overline{\psi}{\not\!\! X}\textrm{i}\gamma) \textrm{D}(\vartheta\psi)+\textrm{D}(\overline{\psi}\vartheta) (\textrm{i}\gamma {\not\!\! X}\psi) \nonumber \\
&&\qquad\qquad\qquad\quad\ \ +(\overline{\psi}\vartheta)\wedge  \textrm{D}(\textrm{i}\gamma{\not\!\! X}\psi)+\textrm{D}(\overline{\psi}{\not\!\! X}\textrm{i}\gamma)\wedge (\vartheta\psi)] \nonumber \\
&&\qquad\qquad\quad=R^{\alpha\beta}\wedge\eta_{\alpha\beta\kappa}\frac{\textrm{i}}{2}\overline{\psi}(\gamma^\kappa{\not\!\! X}-{\not\!\! X}\gamma^\kappa)\psi -R^\alpha{}_\beta\wedge\vartheta^\beta \textrm{i}\overline{\psi}({\not\!\! X}\gamma_\alpha-\gamma_\alpha{\not\!\! X})\gamma\psi \nonumber \\
&&\qquad\qquad\qquad+\textrm{d}[-(\overline{\psi}{\not\!\! X}\textrm{i}\gamma) \textrm{D}(\vartheta\psi)+\textrm{D}(\overline{\psi}\vartheta) (\textrm{i}\gamma {\not\!\! X}\psi) \nonumber \\
&&\qquad\qquad\qquad\quad\ \ +(\overline{\psi}\vartheta)\wedge  \textrm{D}(\textrm{i}\gamma{\not\!\! X}\psi)+\textrm{D}(\overline{\psi}{\not\!\! X}\textrm{i}\gamma)\wedge (\vartheta\psi)]
\end{eqnarray}
Here we attain
\begin{eqnarray}\label{repetitivepattern2}
&&\vartheta^\lambda\overline{\psi}({\not\!\! X}\textrm{i}\gamma\gamma_{\alpha\beta}\gamma_\lambda+\gamma_\lambda
\gamma_{\alpha\beta}\textrm{i}\gamma{\not\!\! X})\psi \nonumber \\
&&=\eta_{\alpha\beta\kappa}\textrm{i}\overline{\psi}(\gamma^\kappa{\not\!\! X}-{\not\!\! X}\gamma^\kappa)\psi-2\vartheta_\beta \textrm{i}\overline{\psi}({\not\!\! X}\gamma_\alpha-\gamma_\alpha{\not\!\! X})\gamma\psi.
\end{eqnarray}
Comparing (\ref{repetitivepattern2}) with (\ref{repetitivepattern1}), we find that when the interchange between $\gamma_\lambda\gamma_{\alpha\beta}$ and $\gamma_{\alpha\beta}\gamma_\lambda$ in the left-hand side is made, the second term in the right-hand side differs by a sign.
Like $\mathcal{H}_1(\psi,{\not\!\! X})$, under the torsion-free presumption, the second term of the first line in the final step of (\ref{HAs2}) disappears and we attain to the desired Hamiltonian volume term plus a total differential, which we are to investigate now if its superpotential can give an expression for the quasilocal angular momentum.

We have known the boundary 2-form contributes a finite but nonvanishing asymptotical integral from (\ref{fot2}) and (\ref{fot3}). Let us see the asymptotical form of it.
\begin{eqnarray}
&&\qquad[-(\overline{\psi}{\not\!\! X}\textrm{i}\gamma) \textrm{D}(\vartheta\psi)+\textrm{D}(\overline{\psi}\vartheta) (\textrm{i}\gamma {\not\!\! X}\psi)+(\overline{\psi}\vartheta)\wedge\textrm{D}(\textrm{i}\gamma{\not\!\! X}\psi)+\textrm{D}(\overline{\psi}{\not\!\! X}\textrm{i}\gamma)\wedge (\vartheta\psi)] \nonumber \\
&&\qquad=2[-(\overline{\psi}{\not\!\! X}\textrm{i}\gamma) \textrm{D}(\vartheta\psi)+\textrm{D}(\overline{\psi}\vartheta) (\textrm{i}\gamma {\not\!\! X}\psi)] \nonumber \\
&&\qquad=\quad2[-(\overline{\psi}{\not\!\! X}\textrm{i}\gamma) \stackrel{\circ}{\textrm{D}}(\vartheta\psi)+\stackrel{\circ}{\textrm{D}}(\overline{\psi}\vartheta) (\textrm{i}\gamma {\not\!\! X}\psi)]\nonumber \\
&&\qquad\quad\,+2[-(\overline{\psi}{\not\!\! X}\textrm{i}\gamma)\wedge\Delta\omega (\vartheta\psi)+(\overline{\psi}\vartheta)\wedge  \Delta\omega( \textrm{i}\gamma{\not\!\! X}\psi)] \nonumber \\
&&\qquad=\quad2[-(\overline{\psi}{\not\!\! X}\textrm{i}\gamma) \stackrel{\circ}{\textrm{D}}(\vartheta\psi)+\stackrel{\circ}{\textrm{D}}(\overline{\psi}\vartheta) (\textrm{i}\gamma {\not\!\! X}\psi)] \nonumber \\ &&\qquad\quad\,-\frac{1}{2}\Delta\omega^{\alpha\beta}\wedge\vartheta^\lambda\overline{\psi}(
{\not\!\! X}\textrm{i}\gamma\gamma_{\alpha\beta}\gamma_\lambda+\gamma_\lambda\gamma_{\alpha\beta}\textrm{i}\gamma{\not\!\! X})\psi \nonumber \\
&&\qquad=\quad2[-(\overline{\psi}{\not\!\! X}\textrm{i}\gamma) \stackrel{\circ}{\textrm{D}}(\vartheta\psi)+\stackrel{\circ}{\textrm{D}}(\overline{\psi}\vartheta) (\textrm{i}\gamma {\not\!\! X}\psi)] \nonumber \\
&&\qquad\quad\,-\Delta\omega^{\alpha\beta}\wedge\eta_{\alpha\beta\kappa}\frac{\textrm{i}}{2}\overline{\psi}(\gamma^\kappa{\not\!\! X}-{\not\!\! X}\gamma^\kappa)\psi +\Delta\omega^\alpha{}_\beta\wedge\vartheta^\beta\textrm{i}\overline{\psi}({\not\!\! X}\gamma_\alpha-\gamma_\alpha{\not\!\! X})\gamma\psi \nonumber \\
\end{eqnarray}
Here, like $\mathcal{H}_1(\psi,{\not\!\! X})$, we deliberately choose the terms for which pulling out $\stackrel{\circ}{\textrm{D}}$ to become $\textrm{d}$ doesn't hold and again, one can see the result is not changed if the alternative is chosen. This expression is arrived by adding (\ref{adding3}) to the boundary expression of (\ref{HAs2}). In the third step (\ref{repetitivepattern2}) is invoked.
The final line is shared with the boundary 2-form of $\mathcal{H}_1(\psi, {\not\!\! X})$ in (\ref{HAP2}) except for a sign difference in the second term and we know they contribute a finite but nonvanishing integral asymptotically from the previous section. Thus we have the redundant second term of the final line alike. Then we come to check the first line of the final step.
\begin{eqnarray}\label{H2BFP}
&&(\overline{\psi}{\not\!\! X}\textrm{i}\gamma) \stackrel{\circ}{\textrm{D}}(\vartheta\psi)\approx\stackrel{\circ}{\textrm{D}}(\overline{\psi}\vartheta) (\textrm{i}\gamma {\not\!\! X}\psi)\approx \textrm{d}\vartheta^-_2{\not\!\! X}+...
\end{eqnarray}
from the process of (\ref{fot2}). It contributes a nonvanishing asymptotic integral. It is not what we desire. Thus the boundary 2-form of the Hamiltonian can't give the angular momentum quasilocalization because it does not approach the ADM term asymptotically.

Let us vary $\mathcal{H}_2$. Varying each of the two terms of the first line in (\ref{HAs2}) with the application of (\ref{variationrule}), we get
\begin{eqnarray}
&&\qquad\delta\mathcal{H}_2(\psi,{\not\!\! X})\nonumber \\
&&\qquad=2\delta[-\textrm{D}(\overline{\psi}{\not\!\! X}\textrm{i}\gamma)\wedge \textrm{D}(\vartheta\psi)+\textrm{D}(\overline{\psi}\vartheta)\wedge \textrm{D}(\textrm{i}\gamma {\not\!\! X}\psi)] \nonumber \\
&&\qquad=\>\>2\textrm{d}[-\delta(\overline{\psi}{\not\!\! X}\textrm{i}\gamma)\stackrel{\circ}{\textrm{D}}(\vartheta\psi)+\delta(\overline{\psi}\vartheta)\wedge \stackrel{\circ}{\textrm{D}}(\textrm{i}\gamma {\not\!\! X}\psi) \nonumber \\
&&\qquad\qquad\>\>\ +\stackrel{\circ}{\textrm{D}}(\overline{\psi}{\not\!\! X}\textrm{i}\gamma)\wedge\delta(\vartheta\psi)+\stackrel{\circ}{\textrm{D}}(\overline{\psi}\vartheta) \delta(\textrm{i}\gamma {\not\!\! X}\psi)] \nonumber \\
&&\qquad\quad\ -\textrm{d}\{\frac{1}{2}\Delta\omega^{\alpha\beta}\wedge\delta[(\overline{\psi}{\not\!\! X}\textrm{i}\gamma) \gamma_{\alpha\beta}(\vartheta\psi)+(\overline{\psi}\vartheta) \gamma_{\alpha\beta}(\textrm{i}\gamma{\not\!\! X}\psi)]\} \nonumber \\
&&\qquad\quad\ +\frac{1}{2}R^{\alpha\beta}\wedge\delta[(\overline{\psi}{\not\!\! X}\textrm{i}\gamma) \gamma_{\alpha\beta}(\vartheta\psi)+(\overline{\psi}\vartheta) \gamma_{\alpha\beta}(\textrm{i}\gamma{\not\!\! X}\psi)] \nonumber \\
&&\qquad\quad\ +\frac{1}{2}\delta\omega^{\alpha\beta}\wedge \textrm{D}[(\overline{\psi}{\not\!\! X}\textrm{i}\gamma) \gamma_{\alpha\beta}(\vartheta\psi)+(\overline{\psi}\vartheta) \gamma_{\alpha\beta}(\textrm{i}\gamma{\not\!\! X}\psi)] \nonumber \\
&&\qquad=\>\>2\textrm{d}[-\delta(\overline{\psi}{\not\!\! X}\textrm{i}\gamma)\stackrel{\circ}{\textrm{D}}(\vartheta\psi)+\delta(\overline{\psi}\vartheta)\wedge \stackrel{\circ}{\textrm{D}}(\textrm{i}\gamma {\not\!\! X}\psi) \nonumber \\
&&\qquad\qquad\>\>\ +\stackrel{\circ}{\textrm{D}}(\overline{\psi}{\not\!\! X}\textrm{i}\gamma)\wedge\delta(\vartheta\psi)+\stackrel{\circ}{\textrm{D}}(\overline{\psi}\vartheta) \delta(\textrm{i}\gamma {\not\!\! X}\psi)] \nonumber \\
&&\qquad\quad\ -\textrm{d}\{\frac{1}{2}\Delta\omega^{\alpha\beta}\wedge\delta[\vartheta^\lambda\overline{\psi}({\not\!\! X}\textrm{i}\gamma\gamma_{\alpha\beta}\gamma_\lambda
+\gamma_\lambda\gamma_{\alpha\beta}\textrm{i}\gamma{\not\!\! X})\psi]\} \nonumber \\
&&\qquad\quad\ +\frac{1}{2}R^{\alpha\beta}\wedge\delta[\vartheta^\lambda\overline{\psi}({\not\!\! X}\textrm{i}\gamma\gamma_{\alpha\beta}\gamma_\lambda
+\gamma_\lambda\gamma_{\alpha\beta}\textrm{i}\gamma{\not\!\! X})\psi]\nonumber \\
&&\qquad\quad\ +\frac{1}{2}\delta\omega^{\alpha\beta}\wedge \textrm{D}[\vartheta^\lambda\overline{\psi}({\not\!\! X}\textrm{i}\gamma\gamma_{\alpha\beta}\gamma_\lambda
+\gamma_\lambda\gamma_{\alpha\beta}\textrm{i}\gamma{\not\!\! X})\psi]\nonumber \\
&&\qquad=\>\> 2\textrm{d}[-\delta(\overline{\psi}{\not\!\! X}\textrm{i}\gamma)\stackrel{\circ}{\textrm{D}}(\vartheta\psi)+\delta(\overline{\psi}\vartheta)\wedge \stackrel{\circ}{\textrm{D}}(\textrm{i}\gamma {\not\!\! X}\psi) \nonumber \\
&&\qquad\qquad\>\>\ +\stackrel{\circ}{\textrm{D}}(\overline{\psi}{\not\!\! X}\textrm{i}\gamma)\wedge\delta(\vartheta\psi)+\stackrel{\circ}{\textrm{D}}(\overline{\psi}\vartheta) \delta(\textrm{i}\gamma {\not\!\! X}\psi)] \nonumber \\
&&\qquad\quad\ -\textrm{d}\{\Delta\omega^{\alpha\beta}\wedge\delta[\eta_{\alpha\beta\kappa}\frac{\textrm{i}}{2}\overline{\psi}(\gamma^\kappa{\not\!\! X}-{\not\!\! X}\gamma^\kappa)\psi] -\Delta\omega^\alpha{}_\beta\wedge\delta [\vartheta^\beta\textrm{i}\overline{\psi}({\not\!\! X}\gamma_\alpha-\gamma_\alpha{\not\!\! X})\gamma\psi]\}
\nonumber \\
&&\qquad\quad\ +R^{\alpha\beta}\wedge\delta[\eta_{\alpha\beta\kappa}\frac{\textrm{i}}{2}\overline{\psi}(\gamma^\kappa{\not\!\! X}-{\not\!\! X}\gamma^\kappa)\psi] -R^\alpha{}_\beta\wedge\delta [\vartheta^\beta\textrm{i}\overline{\psi}({\not\!\! X}\gamma_\alpha-\gamma_\alpha{\not\!\! X})\gamma\psi] \nonumber \\
&&\qquad\quad\ +\delta\omega^{\alpha\beta}\wedge \textrm{D}[\eta_{\alpha\beta\kappa}\frac{\textrm{i}}{2}\overline{\psi}(\gamma^\kappa{\not\!\! X}-{\not\!\! X}\gamma^\kappa)\psi] -\delta\omega^\alpha{}_\beta\wedge \textrm{D}[\vartheta^\beta\textrm{i}\overline{\psi}({\not\!\! X}\gamma_\alpha-\gamma_\alpha{\not\!\! X})\gamma\psi]. \nonumber \\
\end{eqnarray}
In the third step (\ref{repetitivepattern2}) is invoked three times. We see that $\delta\mathcal{H}_2$ has the terms in common with $\delta\mathcal{H}_1$, namely the third, fourth and fifth lines of the final step except for a sign difference for the second term inside the braces of the third line and the second terms of the fourth and fifth lines. So the third line contributes a vanishing asymptotic boundary integral, as checked in the previous section. Let us check the first and second lines of the final step.
\begin{eqnarray}
&&\qquad x^\nu\delta(\overline{\psi}\gamma_\nu \textrm{i}\gamma)\stackrel{\circ}{\textrm{D}}(\vartheta\psi)\approx\stackrel{\circ}{\textrm{D}}(\overline{\psi}\vartheta) x^\nu\delta(\textrm{i}\gamma \gamma_\nu\psi)\approx\textrm{ vanishing integral asymptotically} \nonumber \\
\end{eqnarray}
from (\ref{fpbhv3}).
\begin{eqnarray}
&&\delta(\overline{\psi}\vartheta)\wedge \stackrel{\circ}{\textrm{D}}(\textrm{i}\gamma {\not\!\! X}\psi)\approx\stackrel{\circ}{\textrm{D}}(\overline{\psi}{\not\!\! X}\textrm{i}\gamma)\wedge\delta(\vartheta\psi)
\approx \textrm{d}{\not\!\! X}\delta\vartheta^-_2+...
\end{eqnarray}
from (\ref{fpbhv4}).
The asymptotical boundary integral of the Hamiltonian variation involved with the dynamical variable variation doesn't vanish, so the field equations are not well defined.

In conclusion, $\mathcal{H}_2(\psi, {\not\!\! X})$ is not successful to be a description for the angular momentum due to two reasons: first, even though we choose even parity for the asymptotic $\rm{O}(\frac{1}{r^2})$ term in the spinor field to kill off the undesired terms due to $\psi^-_2$, its boundary 2-form contains other redundant nonvanishing asymptotic terms resulting from $\vartheta^-_2$ and $(\omega^+_3 {\not\!\! X})^-_2$, the latter resulting from the former and being the factor leading the ADM term to contribute a nonvanishing asymptotic integral value, so its boundary 2-form doesn't approach the desired ADM term asymptotically; second, the field equations are not well defined for the part of the boundary 2-form of its variation associated with the dynamical variable variation doesn't have a vanishing asymptotic integral because, again, $\vartheta^-_2$ causes the problem.

\section{A structural analysis of the Witten spinor Hamiltonians adapted for angular momentum}\label{StructuralAnalysis}

The Witten Hamiltonian with the spinor bivectorial parameterization looks far more complicated than the original Witten Hamiltonian with the spinor vectorial parameterization. Actually the variant version of the Witten Hamiltonian bears the structure similar to that of the Quadratic Spinor Lagrangian (QSL) Hamiltonian.

Among the three latter investigated Hamiltonians, $\mathcal{H}_2(\psi, {\not\!\! X})$ is structurally similar to the energy function derived directly from the QSL,
\begin{equation}\label{Eqs1}
\mathcal{E}_{\scriptsize \textrm{qs}1}({\not\!\! N})=2[\textrm{D}(\overline{\psi}{\not\!\! N})\wedge\gamma \textrm{D}(\vartheta\psi)+\textrm{D}(\overline{\psi}\vartheta)\gamma\wedge \textrm{D}({\not\!\! N}\psi)].
\end{equation}
Replacing ${\not\!\! N}=N^\mu\gamma_\mu$ with ${\not\!\! X}$, and based on our need of the spinor bivectorial parameterization, the sign of either of the two terms has to be changed.
Accordingly, we arrive at $2[-\textrm{D}(\overline{\psi}{\not\!\! X})\wedge\gamma \textrm{D}(\vartheta\psi)+\textrm{D}(\overline{\psi}\vartheta)\gamma\wedge \textrm{D}({\not\!\! X}\psi)]$. But we require a real Hamiltonian, which can be realized by two terms complex conjugate to each other. Then we see $\textrm{D}(\overline{\psi}{\not\!\! X})\wedge\gamma \textrm{D}(\vartheta\psi)$ is complex conjugate to $\textrm{D}(\overline{\psi}\vartheta)\gamma\wedge \textrm{D}({\not\!\! X}\psi)$. Hence $\textrm{i}=\sqrt{-1}$ needs to be multiplied for rescue, and thus $\mathcal{H}_2(\psi, {\not\!\! X})$ is achieved. Further, if (\ref{Eqs1}) is inserted with the vanishing torsion constraint, it is equal to
\begin{equation}\label{alternativeEqs1}
2[-\textrm{D}(\overline{\psi}{\not\!\! N}\gamma\vartheta)\wedge \textrm{D}\psi-\textrm{D}\overline{\psi}\wedge \textrm{D}(\vartheta\gamma{\not\!\! N}\psi)],
\end{equation}
which can be modified by the same procedure to get $\mathcal{H}_1(\psi, {\not\!\! X})$. Still further, if we add half (\ref{Eqs1}) and half (\ref{alternativeEqs1}), we attain to the antecedent of $\mathcal{H}(\psi, {\not\!\! X})$.

\section{Conclusion}\label{Conclusion}

Witten and Nester \cite{Witten,PositiveEnergy} constructed a quadratic spinor Hamiltonian expression which provides the gravitational energy-momentum quasilocalization. The Witten-Nester spinor Hamiltonian expression distinguishes from other quasilocal expressions because it leads to a proof of the positive quasilocation of the gravitational energy \cite{Anotherpositivity}. However, whether a spinor Hamiltonian composed by this approach can provide the quasilocalization of the angular momentum and center-of-mass moment kept not being explored. A complete theory of the spinor Hamiltonian formulation is expected to be capable of describing the total 10 conservative quantities of an asymptotically flat gravitational system. The purpose of this work is just to study Witten's approach for the angular momentum.

The usual GR covariant Hamiltonian formalism includes the ``preferred'' boundary term (\ref{B}) with the covariant symplectic structure. With the Regge-Teitelboim (RT) conditions of falloffs and parities, the following properties are fulfilled:
\begin{enumerate}
  \item \label{re1} The Hamiltonian gives the correct dynamical evolution equations and its value is finite.
  \item \label{re2} The part of the boundary integral of the Hamiltonian variation involved with the dynamical variable variation vanishes at spatial infinity.
  \item \label{re3} At spatial infinity the boundary integral yields the desired 10 conserved energy-momentum and angular-momentum/center-of-mass values.
  \item \label{re4} Moreover, at future null infinity (where the fall-off is slower) (a) the preferred boundary integral gives the Bondi energy and (b) the boundary integral of the Hamiltonian variation gives the Bondi energy flux.
\end{enumerate}
Then we should note that the RT falloff-and-parity conditions are sufficient for good values in these limits but are not necessary. One can weaken them considerably. Our approach is to identify a ``good'' boundary term, then one could look for the weakest asymptotic conditions for which the above listed properties hold.

Now maybe our ``preferred'' boundary term could be replaced by one which is even better, such as perhaps some expression using the spinor field. A candidate is the Hamiltonian boundary term associated with the Witten positive energy proof. This is in some ways even better for energy-momentum (it guarantees positive energy).

In Witten and Nester's conception, a Hamiltonian comprising a set of favorable quadratic spinor terms is composed. Then through a certain spinor-curvature identity, the set of the quadratic spinor terms can be transformed into the curvature-involved terms and a total differential. While the curvature-involved terms can, after decomposition and recombination by means of some identity associated with the Dirac algebra, become the standard Hamiltonian volume term under the \textit{a priori} torsion-free condition with the displacement expressed in terms of a specific spinor parameterization, the superpotential of the total differential is expected to yield the quasilocalization of the conservative quantities corresponding to that spinor parameterization of displacement. In the original Witten-Nester Hamiltonian for the energy-momentum, the displacement is expressed in terms of the spinor vectorial parameterization.

But can it be extended to include the angular momentum? In considering alternative boundary terms, our basic test is that a good expression should behave at least as well as the ``preferred'' boundary term in having properties \ref{re1}, \ref{re2}, and \ref{re3} at spatial infinity for the RT asymptotics. This is our minimum requirement. Of course we expect that it should behave well even under weaker than the RT conditions.

For the angular momentum, we need a spinor parameterization which characterizes a Poincar\'{e} rotation for the displacement.
We examined four spinor Hamiltonians in such parameterizations with possible field perturbations. First we substitute $\textrm{i}\overline{\psi}\gamma^\mu\gamma\psi$, the spinor pseudovectorial parameterization, for $\overline{\psi}\gamma^\mu\psi$ in the original Witten Hamiltonian, arriving at $\mathcal{H}_{{\rm{wa}}}(\psi):=2[\textrm{i}\textrm{D}\overline{\psi}\wedge \textrm{D}(\vartheta\psi)-\textrm{i}\textrm{D}(\overline{\psi}\vartheta)\wedge \textrm{D}\psi]$. However, at the very beginning of the investigation, we find this simple idea is problematical because the rotational displacement $\varepsilon^\mu{}_\nu x^\nu$ is a vector instead of a pseudovector. And, going back to the original Witten Hamiltonian with the spinor vectorial parameterization for the rotational displacement would make the Hamiltonian integral divergent, that is, \ref{re1} is not satisfied. Then we try to only express the rotational parameters $\varepsilon^\mu{}_\nu$ as the spinor bivectorial parameterization, conceiving the 4-term quadratic spinor Hamiltonian $\mathcal{H}(\psi,{\not\!\! X}):=-\textrm{D}\overline{\psi}\wedge
\textrm{D}(\vartheta \textrm{i}\gamma {\not\!\! X}\psi)+\textrm{D}(\overline{\psi}{\not\!\! X} \textrm{i}\gamma\vartheta)\wedge \textrm{D}\psi -\textrm{D}(\overline{\psi}{\not\!\! X}\textrm{i}\gamma)\wedge \textrm{D}(\vartheta\psi)+\textrm{D}(\overline{\psi}\vartheta)\wedge \textrm{D}(\textrm{i}\gamma{\not\!\! X}\psi)$ with the parameterization $N^\kappa=\frac{\textrm{i}}{2}\overline{\psi}[\gamma^\kappa{\not\!\! X}-{\not\!\! X}\gamma^\kappa]\psi$. After examining it with a possible choice of the field perturbations, we found it is not successful because it doesn't satisfy \ref{re2} and \ref{re3}. The problems are caused by $\vartheta^-_2$ and $\psi^-_2$. Pinpointing the culprit of the problems, we turned to choose the Hamiltonian consisting of twice the promising two terms among the original 4 terms with the asymptotic spinor behaviour $\psi^-_2$ changed into $\psi^+_2$, $\mathcal{H}_1(\psi, {\not\!\! X}):=
2[-\textrm{D}\overline{\psi}\wedge \textrm{D}(\vartheta \textrm{i}\gamma {\not\!\! X}\psi)+\textrm{D}(\overline{\psi}{\not\!\! X}\textrm{i}\gamma\vartheta)\wedge \textrm{D}\psi]$, which has the same spinor parameterization for the displacement as $\mathcal{H}(\psi,{\not\!\! X})$. We found this Hamiltonian is free of all the problems in $\mathcal{H}(\psi, {\not\!\! X})$, but its boundary form contains a redundant asymptotical term $\Delta\omega^\alpha{}_\beta\wedge\vartheta^\beta \textrm{i}\overline{\psi}(\gamma_\alpha{\not\!\! X}
-{\not\!\! X}\gamma_\alpha)\gamma\psi$, which does not appear in the boundary 2-form of $\mathcal{H}(\psi,{\not\!\! X})$. That's to say, it fails to satisfy \ref{re3}. The culprit is $(\omega^+_3 {\not\!\! X})^-_2$. We also made an inspection of the Hamiltonian consisting of twice the other 2 terms with the choice of the asymptotic field behaviour the same as that for $\mathcal{H}_1(\psi, {\not\!\! X})$, $\mathcal{H}_2(\psi, {\not\!\! X}):=
2[-\textrm{D}(\overline{\psi}{\not\!\! X}\textrm{i}\gamma)\wedge \textrm{D}(\vartheta\psi)+\textrm{D}(\overline{\psi}\vartheta)\wedge \textrm{D}(\textrm{i}\gamma {\not\!\! X}\psi)]$, which has yet the same spinor parameterization for the displacement as $\mathcal{H}(\psi,{\not\!\! X})$, and found it not only has
partial problems of $\mathcal{H}(\psi,{\not\!\! X})$ but also contains the redundant asymptotical boundary term $\Delta\omega^\alpha{}_\beta\wedge\vartheta^\beta \textrm{i}\overline{\psi}({\not\!\! X}\gamma_\alpha-\gamma_\alpha{\not\!\! X})\gamma\psi$, which is the minus of the redundant asymptotical boundary term in $\mathcal{H}_1(\psi, {\not\!\! X})$; it doesn't satisfy \ref{re2} and \ref{re3}. The culprits of the two problems are $\vartheta^-_2$ and $(\omega^+_3 {\not\!\! X})^-_2$. But we can't choose $\vartheta^+_2$ because it would lead the ADM integral to vanish and thus render the vanishment of the total conservative angular momentum, being trivial and meaningless. Therefore neither $\mathcal{H}_1(\psi, {\not\!\! X})$ nor $\mathcal{H}_2(\psi, {\not\!\! X})$ can succeed to be a description for the angular momentum. These spinor Hamiltonians all have the desired 3-volume term under the \textit{a priori} vanishing torsion condition, but if \ref{re2} is not satisfied, the evolution equations are not well defined.

It turns out that we did not succeed in finding any spinor expression that behaves well for the angular momentum even with the conservative RT asymptotics. All the spinor expressions we tried were worse than the ``preferred'' expression. With weaker falloff-and-parity conditions these expressions would be even less satisfactory.

Further efforts need to be made to achieve a Hamiltonian for the angular momentum by Witten's approach. In the proof of the gravitational positive energy by the Witten Hamiltonian, the spinor on the boundary is constrained to satisfy the Witten equation. Perhaps some favorable constraint needs to be made for the spinor on the boundary for the Witten Hamiltonian adapted for the angular momentum or some gauge field can be introduced to eliminate the undesired asymptotical boundary terms and still preserve the desired asymptotical boundary terms. If a successful spinor Hamiltonian for the angular momentum is able to be achieved, we will advance one step further to build the expression for the quasilocal center-of-mass moment. The purported way is to include ${\tilde \textrm{D}}_\beta N^\alpha \textrm{D}\eta_\alpha{}^\beta$, which vanishes when the torsion-free constraint is assumed \textit{a priori}, in the volume term of $\mathcal{H}(\psi, {\not\!\! X})$ and add $-{\stackrel{\circ}{\tilde\textrm{D}}}{}^{[\beta}{\stackrel{\circ}{N}}{}^{\alpha]}\Delta\eta_{\alpha\beta}$ to its boundary term, with the spinor parameterization $\overline{\psi}\textrm{i}\gamma^\kappa{}_\mu\psi$ for the rotation parameters $\varepsilon^\kappa{}_\mu$ in $N^\kappa$. Probably we need to include $\textrm{i}\overline{\psi}\gamma^{\alpha\beta}\psi \textrm{D}\eta_{\alpha\beta}$ in its volume term and add $-\textrm{i}\overline{\psi}\gamma^{\alpha\beta}\psi\Delta\eta_{\alpha\beta}$ to its boundary term. And for the two 2-term Hamiltonians, besides the aforementioned operations, we need to find a way to extirpate $-R^\alpha{}_\beta\wedge\vartheta^\beta \textrm{i}\overline{\psi}[\gamma_\alpha{\not\!\! X}-{\not\!\! X}\gamma_\alpha]\gamma\psi$ in $\mathcal{H}_1(\psi, {\not\!\! X})$ and $-R^\alpha{}_\beta\wedge\vartheta^\beta \textrm{i}\overline{\psi}[{\not\!\! X}\gamma_\alpha-\gamma_\alpha{\not\!\! X}]\gamma\psi$ in $\mathcal{H}_2(\psi, {\not\!\! X})$. This difficulty may be as tough as extirpating the redundant terms, $\Delta\omega^\alpha{}_\beta\wedge\vartheta^\beta \textrm{i}\overline{\psi}(\gamma_\alpha{\not\!\! X}
-{\not\!\! X}\gamma_\alpha)\gamma\psi$ and $\Delta\omega^\alpha{}_\beta\wedge\vartheta^\beta \textrm{i}\overline{\psi}({\not\!\! X}\gamma_\alpha-\gamma_\alpha{\not\!\! X})\gamma\psi$, from their boundary 2-forms.

In addition, an even more exciting and more significant task pursuant to this seminal work is to quest for the connection between the energy (or mass in the convention the velocity of light $c=1$) $m$ and angular momentum $J$ of the gravitational field. The connection is implied to be something like the inequality $m\geq\sqrt{\mid J\mid}$ \cite{Dain2006,Dain2006b,Dain2008,numerical2009,Dain2012,ExtensionmJInequalityMaximal,KeyElementmJInequality,ConvexitymJInequality,DeformationCha,U(1)invariant}. Recalling that the Witten Hamiltonian for the energy-momentum contributes an important result, the proof of the positive energy in the gravitational field, we expect that a successful expression for the quasilocal angular momentum by Witten's approach will contribute a proof of such an inequality. If such a connection can be proved successfully in more general circumstances, that will be a great progress in the field of gravitational theories because it will provide a further norm for the quasilocal expression of the gravitational angular momentum, which has not had sufficiently confining criteria.

\end{document}